\pdfoutput=1
\documentclass[11pt]{article}

\usepackage[a4paper,top=2cm, bottom=2cm, left=2cm, right=2cm]{geometry}
\usepackage[bookmarks=false]{hyperref}

\RequirePackage[OT1]{fontenc}
\RequirePackage{amsthm,amsmath}
\RequirePackage{natbib}
\usepackage{graphicx}
\usepackage{enumerate}

\usepackage{amsmath,amssymb,latexsym}
\usepackage{natbib}
\usepackage{times}
\usepackage{bm}
\usepackage{mathrsfs}           % for marina's special characters

\usepackage{verbatim}
\usepackage{changebar}
\usepackage{stfloats}	
\usepackage{color}
\usepackage{xcolor}

\def\S1{{\mathscr S}}
\def\M1{{\mathscr M}}

\def\beq{\begin{equation}}
\def\eeq{\end{equation}}
\def\beqa{\begin{eqnarray}}
\def\eeqa{\end{eqnarray}}
\def\beqann{\begin{eqnarray*}}
\def\eeqann{\end{eqnarray*}}
\def\bcb{\begin{changebar}}
\def\ecb{\end{changebar}}

\begin{document}

%\begin{frontmatter}

\title{Clustering Nonstationary Circadian Plant Rhythms using Locally Stationary Wavelet Representations}
%\runtitle{Clustering Nonstationary Time Series}
%\thankstext{T1}{Footnote to the title with the ``thankstext'' command.}

\author{Jessica K. Hargreaves\thanks{Corresponding author: {\tt jkh516@york.ac.uk}}\hspace{.2cm}\\
    Department of Mathematics, University of York\\
    Marina I. Knight \hspace{.2cm}\\
    Department of Mathematics, University of York\\
    Jon W. Pitchford\\
    Department of Mathematics and Biology, University of York\\
    and\\
    Seth J. Davis\\
    Department of Biology, University of York}%, Lancaster, LA1 4YF, UK}

\maketitle

\begin{abstract}
How does soil pollution affect a plant's circadian clock? Are there any differences between how the clock reacts when exposed to different concentrations of elements of the periodic table? If so, can we characterise these differences?

We approach these questions by analysing and modelling circadian plant data, where the levels of expression of a luciferase reporter gene were measured at regular intervals over a number of days after exposure to different concentrations of lithium.

A key aspect of circadian data analysis is to determine whether a time series (derived from experimental data) is `rhythmic' and, if so, to determine the underlying period. However, our dataset displays nonstationary traits such as changes in amplitude, gradual changes in period and phase-shifts.

In this paper, we develop clustering methods using a wavelet transform. Wavelets are chosen as they are ideally suited to identifying discriminant local time and scale features. Furthermore, we propose treating the observed time series as realisations of locally stationary wavelet processes. This allows us to define and estimate the evolutionary wavelet spectrum. We can then compare, in a quantitative way, using a functional principal components analysis, the time-frequency patterns of the time series. Our approach uses a clustering algorithm to group the data according to their time-frequency patterns. We demonstrate the advantages of our methodology over alternative approaches and show that it successfully clusters our data.
\end{abstract}


\section{Introduction}\label{sec:intro}

The earth rotates on its axis every 24 hours producing the cycle of day and night. Correspondingly, almost all species exhibit changes in their behaviour between day and night \citep{bell2005circadian}. These daily rhythms are not simply a response to the changes in the physical environment but, instead, arise from a timekeeping system within the organism \citep{vitaterna2001overview}. This timekeeping system, or biological `clock', allows the organism to anticipate and prepare for the changes in the physical environment that are associated with day and night. Indeed, most organisms do not simply respond to sunrise but, rather, anticipate the dawn and adjust their biology accordingly. For example, when deprived of external time cues, many of these diurnal rhythms persist, indicating they are maintained by a biological circadian clock within the organism \citep{mcclung2006plant}. The mechanisms underlying the biological timekeeping systems and the potential consequences of their failure are among the issues addressed by researchers in the field of circadian biology.

\subsection{The History of Clock Research in Plants}
\label{sec:history}

\begin{figure}
\centering
\includegraphics[width=0.7\linewidth]{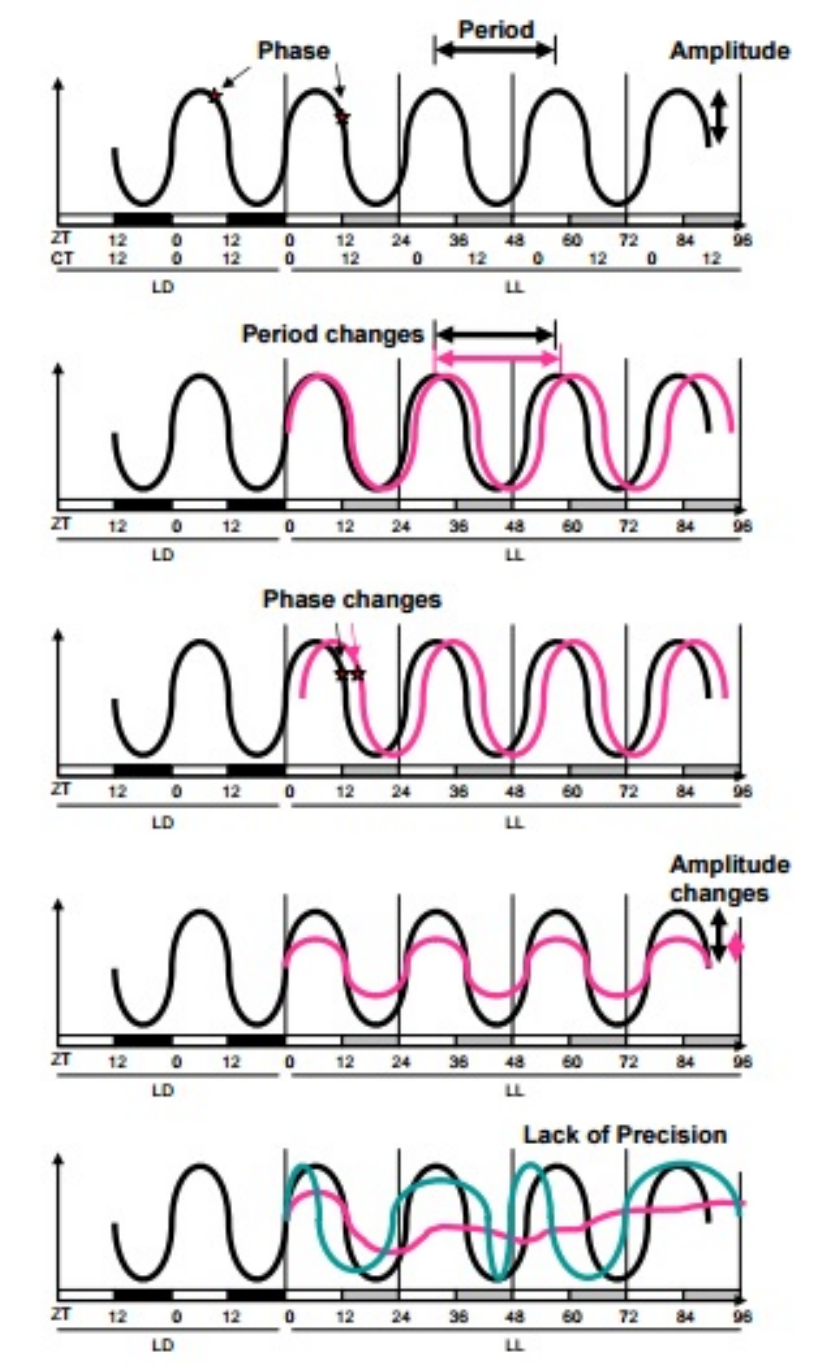}
\caption{The defined rhythmic parameters: periodicity, phase, amplitude and clock precision (taken from \cite{hanano2006multiple}).}
\label{fig:s1}
\end{figure}

Circadian rhythms are the subset of biological rhythms with period, defined as the time to complete one cycle, of approximately 24 hours (see Figure \ref{fig:s1} for a visual interpretation of this terminology). A second defining attribute of circadian rhythms is that they are `endogenously generated and self-sustaining' \citep{mcclung2006plant}. In particular, the period remains approximately 24 hours under constant environmental conditions, such as constant light (or dark) and constant temperature (i.e. when deprived of any external time cues).

\cite{millar1991circadian} identified a number of genes that were under circadian control in a particular plant, the Arabidopsis thaliana. However, these initial Arabidopsis experiments were extremely labour intensive, as plants needed to be individually processed by a researcher at regular intervals over a sustained period of time. One solution was a firefly luciferase system. This method fuses the gene of interest to a luciferase reporter gene. Thus, when the gene is expressed, so is luciferase and the plant produces light. A machine then measures bioluminescence (light emitted from the plant) at regular intervals to obtain a measure of the amount of the gene expressed at a given time \citep{southern2005circadian}.

 \subsection{Current Methods}

 As discussed in Section \ref{sec:history}, in circadian biology, a key aspect of data analysis is to estimate the period of the time series. A current approach in the circadian community would be to use the computer programme BRASS to obtain a period estimate for each time series using Fourier analysis. (See \cite{moore2014online} for a complete description of these methods.) These methods require an underlying assumption that the data is stationary. However, nonstationary behaviour is common in biologocial systems \citep{moore2014online} and our particular dataset displays nonstationary behaviour. Therefore, we propose methods that are capable of detecting \textit{changes of period over time}.

\subsection{Aim and Structure of the Paper}

In this project we wish to understand how a plant's clock is effected when exposed to different concentrations of lithium. Furthermore, we would also like to know which concentrations produce similar effects and then characterise these effects. These questions have important implications for understanding the
mechanism of the plant's circadian clock and also environmental implications associated with soil pollution. Therefore, we develop methods to group and characterise our circadian dataset which comprises of gene expression levels measured at regular time intervals.

In this paper, we demonstrate the time series in our circadian dataset are nonstationary. Therefore, we argue that the current methods used by the circadian community to analyse such data (which assume stationarity) are inappropriate. Therefore, we develop clustering methods using a wavelet transform. Wavelets are chosen as they are ideally suited to identifying discriminant local time and scale features. Furthermore, we propose treating the observed time series as realisations of locally stationary wavelet (LSW) processes. This allows us to define and rigorously estimate the evolutionary wavelet spectrum. We can then compare (in a quantitative way) using a functional principal components analysis, the time-frequency patterns of the time series. Our approach uses a clustering algorithm to group the data according to their time-frequency patterns. Our LSW-based approach combines the use of wavelets with (rigorous stochastic) nonstationary time series modelling and achieves superior results to current methods of clustering nonstationary time series. Furthermore, our method can also be used to produce visualisations that can be used to characterise the features associated with a particular cluster.

This article is organized as follows. In Section \ref{sec:data}, we outline the circadian dataset and the pre-processing that we apply to these data. We also perform a hypothesis test of (second-order) stationarity. In Section \ref{sec:model} we develop a locally stationary wavelet clustering method and describe its implementation. The results of clustering the circadian dataset using the proposed methodology are presented in Section \ref{sec:app}. We also examine the results of our analysis in the context of several relevant biological questions in Section \ref{sec:app} before concluding with a brief discussion in Section \ref{sec:concs}.

\section{Data and Preliminary Processing}\label{sec:data}

In this section we outline the dataset of circadian plant rhythms and the pre-processing that we apply to these data. We also report the results of the classical analysis a circadian biologist would use to analyse such data. Finally, we describe the features of the data which motivate our proposed methods. In particular, we perform a hypothesis test for second-order stationarity.

\begin{figure}
\centering
\includegraphics[width=\linewidth]{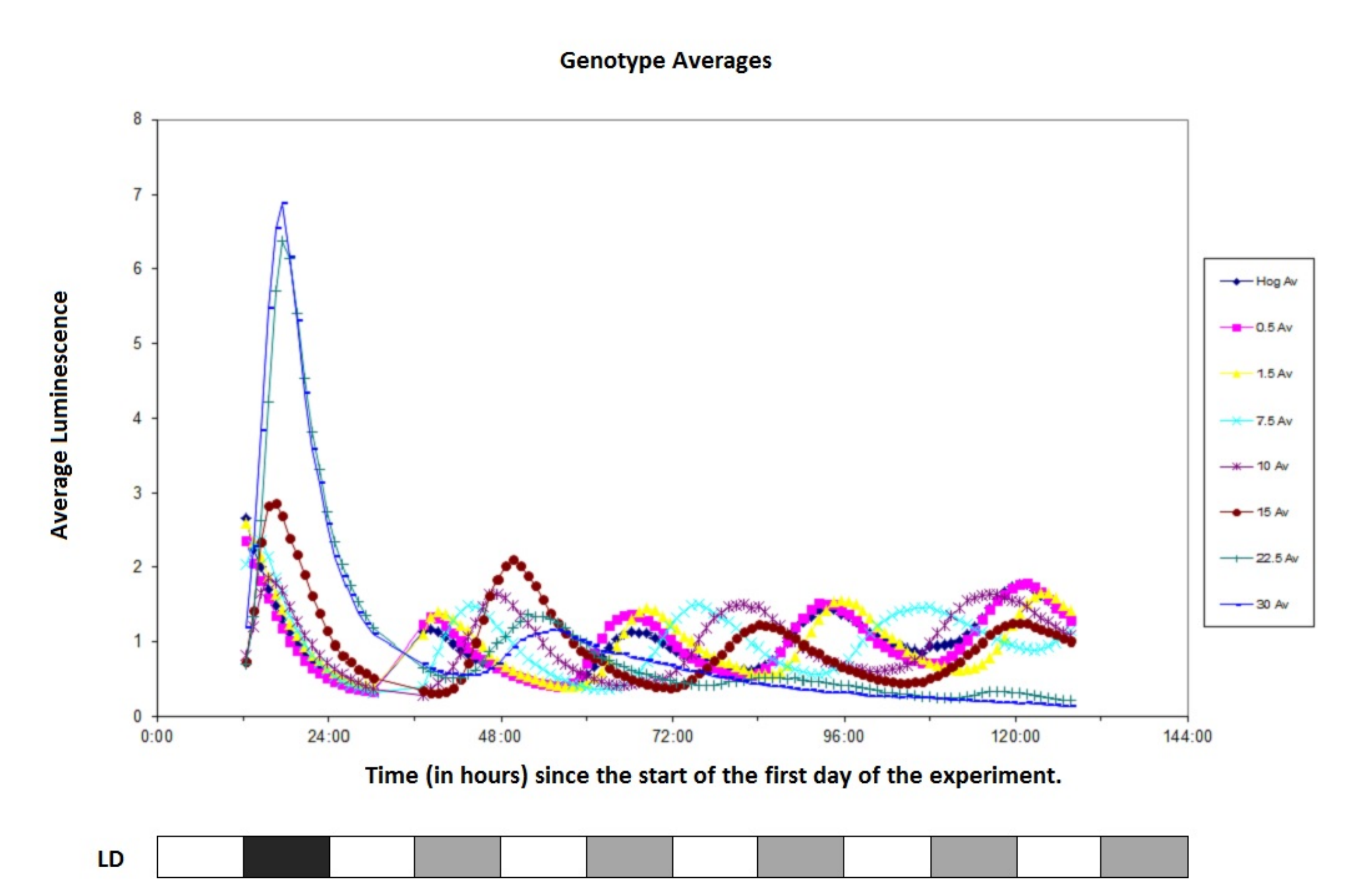}
\caption{A plot of the average expression at each time point for each of the 8 groups of the circadian dataset. Time is measured in hours relative to zeitgeber time, which is the time of the last external temporal cue: the dawn signal of lights-on.}
\label{fig:avplot}
\end{figure}

To obtain this data set, the lab uses a firefly luciferase system. This method fuses the gene of interest, in this experiment CCR2, to a luciferase reporter gene. Thus, when the gene is expressed, so is luciferase and the plant (in this experiment, the Arabidopsis thaliana) emits light. The researcher then measures bioluminescence (light emitted from the plant) to obtain a measure of the amount of the gene expressed \citep{southern2005circadian}. The luminescence rhythms were monitored using a luminescence scintillation counter, TOPCount NXT (Perkin Elmer). This method allows for a quantitative, real-time gene expression measure in living plants. (See, for example, \citep{perea2015modulation} for a complete description of this type of experiment.) A plot of the average expression at each time point for each of the groups is shown in Figure \ref{fig:avplot} (on page \pageref{fig:avplot}). Note that time is measured in hours relative to zeitgeber time, which is the time of the last external temporal cue: the dawn signal of lights-on.

Prior to the recordings in Figure \ref{fig:avplot}, the plants are grown under 12 hour light/12 hour dark cycles to simulate a `normal' day. The plants are then transferred to the TOPCount machine. In this experiment, measurements were taken at equal intervals of approximately 1 hour.  Measurement began after the transition to 12 hours of darkness on a given day. After this, the plant was exposed to constant light. In Figure \ref{fig:avplot}, we can see average luminescence during a "normal" day of twelve hours of light followed by twelve hours of darkness, before exposure to constant light (for approximately 4 days). This is represented by the shaded bars below the graph (the plants are under constant light throughout the experiment, other than between 12:00 and 24:00 hours when the plants are in darkness. This is indicated by the black bar. The grey bars indicate that, though the plants were in constant light, they would be in darkness during a `normal' 12 hour light/12 hour dark cycle.)

Our data set consists of 96 time series recorded at 106 time points. The 96 time series are known to be derived from adding 8 different concentrations of lithium to the plants. In particular, a control group is grown in Hoagland media \citep{hoagland1950water} which contains essential nutrients required for plant growth and is not exposed to any additional levels of lithium. The other seven groups are also grown in the Hoagland media with varying additional concentrations of lithium. In particular, 0.5mM, 1.5mM, 7.5mM, 10mM, 15mM, 22.5mM and 30mM of lithium chloride (LiCl) were respectively added to the Hoagland media to obtain the 7 additional groups. Thus, the data consists of 8 groups of 12 plants with plants from the same group having been treated in the same way.

\begin{figure}
\centering
\includegraphics[width=\linewidth]{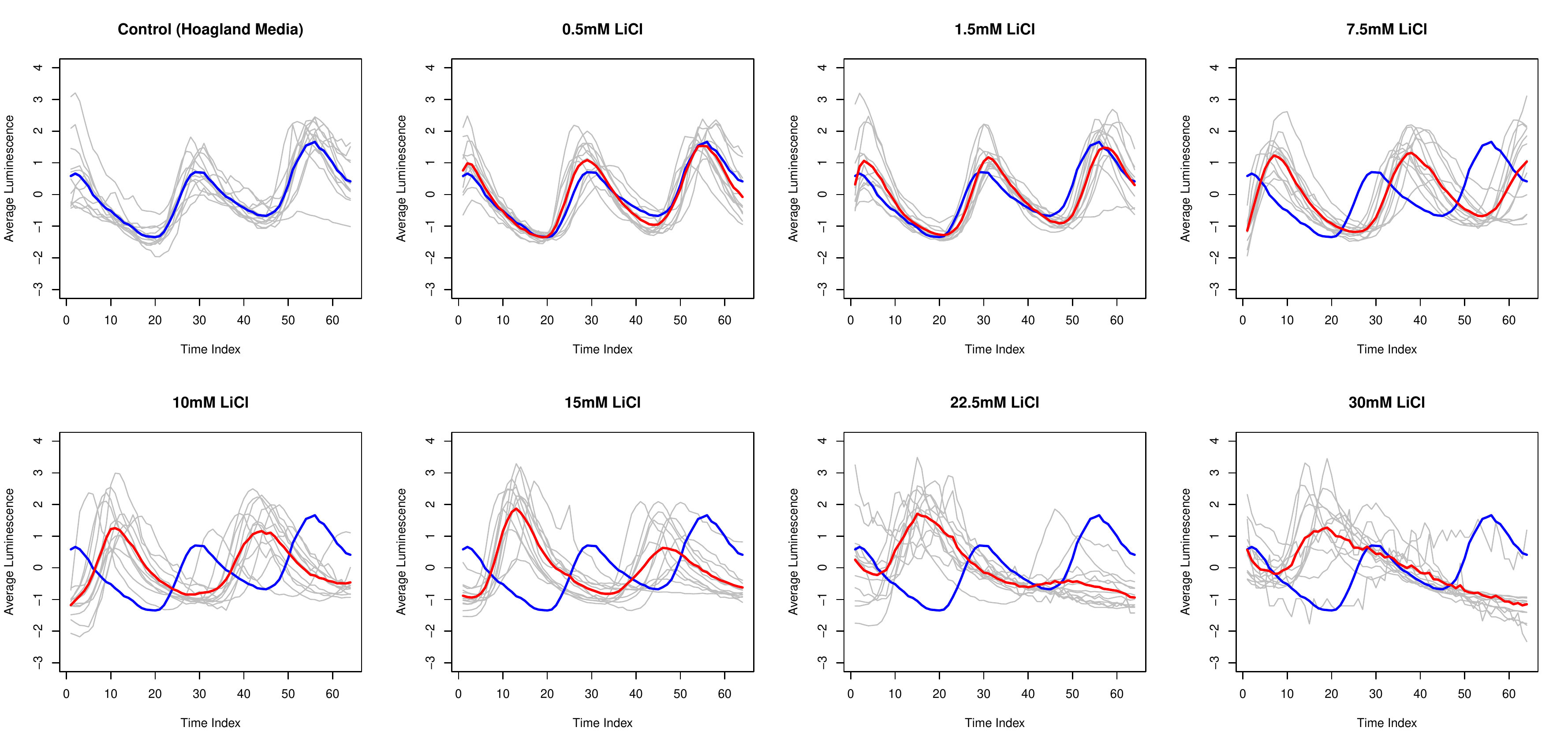}
\caption{Top left: Each realisation from the control group (in grey) along with the group average (in blue). Other panels: Each realisation from each group (in grey) along with the group average (in red) and the control group average (in blue) for our truncated circadian dataset. (Each time series has been normalised to have mean zero.)}
\label{fig:Circgroups}
\end{figure}

On examining Figure \ref{fig:avplot}, we notice a break in the data at time 30:03 for approximately 7 hours. One of the disadvantages of using the system described above to measure gene expression is the propensity of the recording equipment to break down resulting in gaps in the data. For the purposes of this analysis, we will truncate our dataset and examine only the observations after this point. However, we will discuss alternative methods to overcome this problem in Section \ref{sec:concs} as avenues of further work. Under the LSW modelling framework, we require our data to be of length $N=2^J$ (see Section \ref{sec:model}). Each plant was observed at 106 time points; therefore, we chose to truncate the data and use the 64 observations after the break (from time 37:12 to 103:46 in Figure \ref{fig:avplot}). We use these particular observations as they display changes to the `rhythmic parameters' outlined in Section \ref{sec:data description} which are of interest to the circadian biologist such as: period, phase, amplitude and precision. The truncated dataset of $64$ observations can be seen in Figure \ref{fig:Circgroups}.

\subsection{Data Description}
\label{sec:data description}
On examining Figures \ref{fig:avplot} and \ref{fig:Circgroups}, there is visual evidence to suggest that adding lithium produces an effect in the circadian clock of this plant. In particular, the circadian biologist is interested in how certain `rhythmic parameters' of the clock are effected. These defined parameters are periodicity, phase, amplitude and clock precision. Figure \ref{fig:s1} (taken from \cite{hanano2006multiple}) provides graphical description of these parameters.

In particular, there seems to be a pronounced effect after adding 7.5mM (or more) of LiCl whereas the first 3 groups (the control and concentrations 0.5mM and 1.5mM represented by the dark blue, pink and yellow lines in Figure \ref{fig:avplot} or the first three panels in Figure \ref{fig:Circgroups}) are relatively indistinguishable. All concentrations up to and including 15mM seem to display cyclic behaviour whereas concentrations 22.5mM and 30mM do not. In conclusion, these plots indicate that adding increasing concentrations of lithium lengthens the period and also produces a dampening effect.

\subsection{BRASS Results}
In the circadian community, analysis of this data would typically be performed by the Microsoft Excel macro BRASS (see \cite{moore2014online} for a detailed description of this software package and its advantages and disadvantages). Table \ref{tab:BRASS} provides a summary of the output of the analysis the circadian dataset in BRASS. In particular, it shows the mean period estimate for each of the 8 groups and the number of plants that could be analysed by BRASS.

\begin{table}
\begin{tabular}{|p{30mm}|c|c|c|c|c|c|c|c|}
\hline  & Hoagland & 0.5mM & 1.5mM & 7.5mM & 10mM & 15mM & 22.5mM & 30mM \\
\hline Number of plants & 12 & 12 & 12 & 12 & 12 & 12 & 12 & 12 \\
\hline Number of \newline "rhythmic" plants & 11 & 12 & 12 & 12 & 12 & 11 & 3 & 3 \\
\hline Period estimate \newline (in hours) & 26.85 & 26.85 & 26.87 & 30.96 & 31.55 & 27.03 & 18.00 & 22.65 \\
\hline
\end{tabular}
\caption{A summary of the output of the analysis of the circadian dataset in BRASS.}
\label{tab:BRASS}
\end{table}

\medskip

The results of the analysis in BRASS (in Table \ref{tab:BRASS}) show that not all the data is used to produce the period estimate reported by BRASS (the number of `rhythmic' plants is the number of time series which BRASS was able to return a period estimate for). For example, in the 22.5mM and 30mM groups, BRASS was only able to estimate a period for 6 of the 24 plants. This shows that BRASS is not able to analyse all the data produced by this experiment and indicates that this dataset is not suitably modelled using Fourier methods.

\subsection{Test of Stationarity}

As discussed in the previous section, we have reason to believe that this data is nonstationary. In Figure \ref{fig:Circgroups}, the period and amplitude of individual groups seems to change with time. These features indicate that the data is nonstationary and that Fourier analysis is not appropriate for this kind of data. In this section, we investigate whether our truncated dataset is (second-order) stationary by performing a hypothesis test.

The test we will use is based on the methods in \cite{priestley1969test}, the Priestley-Subba Rao (PSR) Test. The PSR test is implemented in the \verb|fractal| package in R available from the CRAN package repository and the results can be found in Table \ref{tab:stattest}. This shows number of (the truncated) time series which did not indicate enough evidence to reject the null hypothesis of stationarity (at the 1\% significance level) for each group. This analysis indicates that very few (approximately 23\%) of the time series did not provide enough evidence to reject the null hypothesis of stationarity. This suggests that the data are nonstationary and the current Fourier analysis methods are not suitable.

\begin{table}
\begin{tabular}{|p{30mm}|c|c|c|c|c|c|c|c|}
\hline  & Hoagland & 0.5mM & 1.5mM & 7.5mM & 10mM & 15mM & 22.5mM & 30mM \\
\hline Number of plants & 12 & 12 & 12 & 12 & 12 & 12 & 12 & 12 \\
\hline Number of \newline "stationary" plants & 4 & 6 & 6 & 2 & 1 & 1 & 0 & 2 \\
\hline
\end{tabular}
\caption{Results for test of stationarity.}
\label{tab:stattest}
\end{table}

\section{Proposed Clustering Method}\label{sec:model}

In their review of period estimation methods for circadian data, \cite{moore2014online} recommend wavelet-based methods for nonstationary time series to extract changes of period over time. In this work, we combine the use of wavelets with (rigorous stochastic) nonstationary time series modelling. Our approach is to cluster the data according to their time-frequency patterns by assuming the locally stationary wavelet model.

\subsection{Modelling Nonstationary Time Series}

  Many approaches to modelling nonstationary time series have been developed from the spectral representation of stationary time series. This states that if a time series $\{X_t\}_{t \in \ \mathbb{Z}}$ is a \emph{stationary} stochastic process, then it admits the following Cram\'er-Rao representation \citep{priestley1983spectral}:

  \begin{equation}
  \label{SRT}
  X_t = \int_{-\pi}^{\pi} A(\omega)\exp(i\omega t ) d\xi(\omega),
  \end{equation}

  where $A(\omega)$ is the amplitude of the process and $d\xi(\omega)$ is an orthonormal increments process.

    \subsubsection{Locally stationary Fourier model}

  In the representation of stationary process in \eqref{SRT}, we note that, for stationary processes, the amplitude $A(\omega)$ does not depend on time (i.e. the frequency behaviour is the same across time). For many real time series, such as our circadian plant data, this is not realistic and the model \eqref{SRT} is inadequate. Therefore, we would prefer a model where the frequency behaviour can vary with time. One way of introducing time dependence into a model is to replace the $A(\omega)$ in \eqref{SRT} with a time dependent form. Therefore, \cite{priestley1965evolutionary} introduced a time-frequency model which is analogous to \eqref{SRT} with the amplitude replaced by $A(\omega, t)$. \cite{dahlhaus1997fitting} extended this and introduced the locally stationary modelling philosophy and developed the locally stationary Fourier (LSF) model. In this setting, the time-dependent transfer function is defined on "rescaled time" to enable asymptotic considerations.

  \subsubsection{Locally stationary wavelet model and assumptions}

  An alternative approach to the LSF model is to replace the Fourier functions in \eqref{SRT} by \textit{wavelets}. The advantage of wavelets is that they are localised in both time and frequency and are therefore well-suited to modelling second-order characteristics that evolve over time. (We refer the reader to \cite{daubechies1992ten} and \cite{nason2008wavelet} for an introduction to wavelets and their application in statistics.) \cite{nason2000wavelet} developed an alternative method  for modelling nonstationary time series by replacing the Fourier functions in \eqref{SRT} by a set of discrete non-decimated wavelets \citep{nason1995stationary}. They proposed the  Locally Stationary Wavelet (LSW) time series model.

  We will now define the LSW model for locally stationary time series as in \cite{nason2000wavelet}. We assume that the reader is familiar with the Discrete Wavelet Transform (for further information see \cite{mallat1989multiresolution} and \cite{mallat1989theory} or \cite{nason2008wavelet} for a complete introduction). We also assume that the reader is familiar with the Non-decimated Wavelet Transform (for a detailed description see \cite{nason1995stationary} and \cite{nason2000wavelet}). Therefore, we define the LSW model as follows.

  The locally stationary wavelet (LSW) process  $\{X_{t;T}\}$ defined to be a sequence of (doubly-indexed) stochastic processes having the following representation in the mean-square sense:

  \begin{equation}
  \label{LSW rep}
  X_{t,T} = \sum_{j = 1}^{\infty} \sum_k w_{j, k;T} \psi_{j,k}(t)\xi_{j,k},
  \end{equation}
  where $\{\xi_{j,k}\}$ is a random orthonormal increment sequence, $\{\psi_{j, k}(t) = \psi_{j, t-k} \}_{j,k}$ is a set of discrete non-decimated wavelets and $\{ w_{j, k;T} \}$ is a set of amplitudes.

  This LSW process formulation also requires three additional assumptions \citep{nason2000wavelet}:

  \begin{enumerate}
    \item $\mathbb{E}(\xi_{j,k}) = 0$ which ensures $X_{t,T}$ is a zero mean process. In practice, if the time series has a non-zero mean, we estimate it and remove it.

    \item $\text{cov}(\xi_{j,k} \xi_{l,m}) = \delta_{j,l}\delta_{k,m}$, where $\delta_{j,l}$ is the Kronecker delta. This ensures the orthogonal increment sequence is uncorrelated.

    \item $\sup_k \lvert w_{j,k;T} - W_j(k/T) \rvert \leq C_j / T,$

  where $W_j(z), z \in (0,1)$ is a function (with various smoothness constraints) and $\{ C_j \}_j$ is a set of constants with $\sum_{j=1}^{\infty} C_j < \infty$.  This condition controls the speed of evolution of $w_{j,k;T}$ and thus permits estimation \citep{nason2008wavelet}.
  \end{enumerate}

   For an extensive review of the locally stationary wavelet model and its application in nonstationary time series analysis, see for example \cite{nason2000wavelet} or \cite{nason2008wavelet}.

  In the stationary time series setting, recall the \textit{spectrum} of a time series, $f(\omega) = |A(\omega)|^2$ where $A(\omega)$ is the amplitude as in equation \eqref{SRT}. The spectrum quantifies the contribution to the variance of a stationary process over a \textit{frequency}, $\omega$. An analogous quantity can be defined for the LSW model \citep{nason2000wavelet}, the evolutionary wavelet spectrum (EWS). The EWS quantifies how the power in an LSW process is distributed across time and scale.

  %\theoremstyle{definition}
  %\begin{defn}
  The evolutionary wavelet spectrum (EWS) of the time series $\{X_{t,T}\}_{t=0}^{T-1}$ is defined as:

  \begin{equation}
  \label{EWS eq}
  S_j(z) = |W_j (z)|^2,
  \end{equation}
  for $j \in \mathbb{N}, z \in (0,1).$
  %\end{defn}
  The quantity $z$, known as \emph{rescaled time}, $z = k/T, z \in (0,1)$, was introduced by \cite{dahlhaus1997fitting} and is used to enable asymptotic considerations (see \cite{fryzlewicz2009consistent} and, for a more detailed discussion, \cite{dahlhaus1996kullback}).

   The (raw) wavelet periodogram is given by:
     \begin{equation}
     \label{periodogram}
     I_{k,T}^j = \lvert d_{j,k;T} \rvert^2,
     \end{equation}
   where
  \begin{equation}
  \label{NDWTds}
  d_{j,k;T} = \sum_{t=1}^T X_{t, T} \psi_{j,k}(t),
  \end{equation}
  are the \textit{empirical non-decimated wavelet coefficients} and it is a biased estimator of the EWS. However, an unbiased \textit{corrected} periodogram may be obtained by premultiplying \citep{nason2000wavelet} the raw wavelet periodogram by the inverse of the autocorrelation wavelet inner product matrix, $A_J$ with

  \[
  A_{j,l} = <\Psi_j, \Psi_l > = \sum_\tau  \Psi_j(\tau)\Psi_l(\tau),
  \]

  and where $\Psi_j (\tau) = \sum_k \psi_{j,k}(0) \psi_{j,k} (\tau)$ for all $ j \in \mathbb{N}$ and $\tau \in \mathbb{Z}$ is the \emph{autocorrelation wavelet} (see \cite{nason2008wavelet} for more information).

  Thus, the \textit{corrected} wavelet periodogram is:
  \begin{equation}
  \label{Corrected periodogram}
  L_{k, T} = A_J^{-1} I_k,
  \end{equation}
  for $k = 0, \dots, T - 1$.

  As in the stationary setting, the wavelet periodogram is not a consistent estimator of the EWS \citep{nason2008wavelet}. One method to overcome this problem is to smooth the wavelet periodogram as a function of (rescaled) time for each scale, $j$. \cite{nason2000wavelet}  recommend smoothing the wavelet periodogram first and then applying the correction as in \eqref{Corrected periodogram} as this is more analytically tractable. In particular, we could smooth the periodogram by log transform. The idea behind using a log transform is that (for Gaussian data) the distribution of the raw wavelet periodogram is approximately $\chi^2$ and the use of the log stabilises the variance and draws the distribution towards normality, thereby permitting universal thresholding, which is designed to work in this situation. Alternative approaches using wavelet-Fisz transforms for smoothing using  variance stabilisation techniques have been proposed by \cite{fryzlewicz2006haar}. We will denote the corrected and smoothed periodogram of time series $\{X_{t,T}\}_{t=0}^{T-1}$ as $\{\hat{S}_{j}(z)\}_{j}$.  For a rigorous discussion of the estimation of the EWS see \cite{nason2000wavelet}.

  Throughout this article, we work with \textit{normal} LSW processes (i.e., the $\xi_{j,k}$ in \eqref{LSW rep} are distributed $N(0,1)$). This is for mathematical convenience and extensions to other distributions are possible. In particular, this results in the wavelet periodogram, $I_{k,T}^j$, having a scaled $\chi_1^2$ distribution \citep{fryzlewicz2009consistent}. As stated above, this assumption allows us to perform universal thresholding. Moreover, under this assumption, the correction of the wavelet periodogram (see Equation \eqref{Corrected periodogram}) brings its distribution closer to Gaussianity \citep{fryzlewicz2009consistent}.

\subsection{Current Clustering/Classification Techniques Taking into Account Nonstationarity}
\label{sec:review}

\cite{shumway2003time} considers the use of time-varying spectra for classification and clustering nonstationary time series. This method uses locally stationary Fourier models and Kullback-Leibler discrimination measures to classify seismic data.
\cite{fryzlewicz2009consistent} develop a procedure for \textit{classification} of nonstationary time series. In their setup, they have available a training dataset that consists of signals with known group labels. The observed signals are viewed as realisations of locally stationary wavelet processes and the evolutionary wavelet spectrum is estimated. The EWS, which contains the second-moment information on the signals, is used as the classification signature. \cite{fryzlewicz2009consistent} thus combine the use of wavelets with rigorous stochastic nonstationary time series modelling. \cite{krzemieniewska2014classification} developed this method in the context of an industrial experiment by proposing an alternative divergence index to compare the spectra of two time series.

Using maximum covariance analysis (MCA) on the wavelet representations of \textit{two} series with clustering applications has been proposed in previous works. MCA is one method to extract common time-frequency patterns and also reduce the dimension of the data. \cite{rouyer2008analysing} use MCA to compare, in a quantitative way, the wavelet spectra of \textit{two} time series. Their approach applies a singular value decomposition on the covariance matrix between each pair of wavelet spectra. The distance between two wavelet spectra is measured by comparing a given number of the leading patterns and singular vectors obtained by the MCA that correspond to a fixed percentage of the total covariance. This is repeated for each pair of time series to build a distance matrix which is used to obtain a cluster tree that groups \textit{wavelet spectra} according to their time-frequency patterns. \cite{antoniadis2013clustering} also use a maximum covariance analysis over a localised covariance matrix, however, their methods are based on the \textit{continuous wavelet transform}. They introduce a way of quantifying the similarity of these patterns by comparing the evolution in time of each pair of leading patterns. This builds a distance matrix which is then used within classical clustering algorithms to differentiate among high dimensional populations.

 \cite{holan2010modeling} proposed achieving dimension reduction by treating a spectrum as an `image' and performing a functional principal components analysis. \cite{holan2010modeling} classify nonstationary time series using a generalised linear model that incorporates the (dimension-reduced) spectrogram of a short-time Fourier transform into the model as a predictor.

\subsection{Proposed Functional PCA Approach for EWS Content}
\label{sec:PCA}

In this section we explain how we will use the features of the corrected spectra to cluster the plants based on their estimated time-scale behaviour. Our approach differs from those outlined in Section \ref{sec:review} as we develop the functional PCA approach for wavelets under the LSW modelling framework. This combines the use of a dimension-reduced wavelet representation with rigorous stochastic nonstationary time series modelling. In particular, we are able to calculate an unbiased, consistent estimator of the EWS and use this as the basis of our clustering methodology.

In our biological problem of interest, the time-frequency representation of the signal is high dimensional. For example, the estimated spectrum of an individual time series from our data set consists of 64 time points by 6 frequencies and thus produces 384 possible time-frequency covariates. Therefore, we need to perform a dimension reduction technique. The method we use treats the spectrum as an "image" and the spectral coefficients as time-frequency "pixels". The pixels are not independent as they result from covariance present in the original signal. In fact, the spectrum presents coherent patterns that should be accounted for.

To address the dependence in the spectrum, \cite{holan2010modeling} treat each spectrum as an image and decompose it as a Karhunen-Lo\' eve representation \citep{james2005functional}. In theory, the spectrum that results from a (continuous) time series is a continuous two-dimensional object. Therefore, consider a continuous spectrum $\{S(\mathbf{v}) : \mathbf{v} = (j, z), \mathbf{v} \in \mathbb{R} \times (0, 1)\}$. Suppose that $E[S(\mathbf{v})]=0$, and define the covariance function as $E[S(\mathbf{v}), S(\mathbf{v}')] \equiv C_S(\mathbf{v}, \mathbf{v}')$. Then the Karhunen-Lo\' eve expansion allows the covariance function to be decomposed via a classical eigenvalue/eigenfunction decomposition.

Although the continuous Karhunen-Lo\' eve representation is often the most realistic from the point of view of modelling a biological process, it is rarely considered in applications. This is due to the discrete nature of observations resulting from most experiments. In practice, we use the empirical version of the Karhunen-Lo\' eve decomposition, the Karhunen-Lo\' eve Transform (also known as empirical orthogonal function (EOF) analysis in meteorology and geophysics) as is common in spatial statistics \citep{cressie2015statistics}.

Therefore, assume we have observed $N$ (LSW) processes at $T = 2^J$ equally spaced time points. Denote the $i$th time series $\{X_{t,T}^{(i)}\}_{t=0}^{T-1}$ for $ i = 1, \dots, N$. For each time series, $\{X_{t,T}^{(i)}\}_{t=0}^{T-1}$, calculate the corrected and smoothed periodogram, $\{\hat{S}^{(i)}_{j}(t/T)\}_{j}$, for $i = 1, \dots, N$, where $t = 0, \dots, T - 1$ and $j = 1, \dots, J$.  The resulting estimated spectral coefficients can be arranged in $N, J \times T$ matrices which we denote $\hat{S}^{(1)}, \dots, \hat{S}^{(N)}$. We can treat each of these matrices as an "image". In particular, for $i = 1, \dots, N$, "vectorise" the matrix $\hat{S}^{(i)}$. That is, concatenate the rows of the matrix $\hat{S}^{(i)}$ to produce a vector $\mathbf{\hat{s}}^{(i)}$ that has length $J \times T = n$. These $N$ vectors are combined to form an $N \times n$ data matrix, $Q$,  where each row of $Q$ represents one matrix, $\hat{S}^{(i)}$. Formally,
 \begin{equation}
 \label{eq:datamat}
 Q = \left[\mathbf{\hat{s}}^{(1)}, \dots, \mathbf{\hat{s}}^{(N)} \right]^T.
 \end{equation}

 This results in a data matrix, $Q$, on which we can perform a classical principal components analysis (PCA) as in multivariate statistics. Therefore, we now briefly outline how to perform a PCA.

 Firstly, note that the data should be centred. Therefore, subtract the column mean from each column of $Q$ and denote the mean centred data matrix $U$. Now compute the sample covariance matrix, $R$, of (the transpose of) the mean centred data matrix, $U^T$:
 \begin{equation}
 R = QQ^T.
 \end{equation}

 Apply the singular value decomposition to the covariance matrix, $R$:
 \begin{equation}
 R = \Psi \Lambda \Psi^T,
 \end{equation}
 where the columns of $\Psi$ are the eigenvectors (known as the \emph{singular vectors}) of $R$ and $\Psi$ is a diagonal matrix whose diagonal elements are the eigenvalues of $R$ (arranged in decreasing order of magnitude), referred to as the \emph{singular  values}. The singular values are proportional to the squared covariance accounted for in each direction.

 We can project the data matrix, $U$, onto its principal components by multiplying: $UU^T \Psi$. Call these projected values the scores of each process.

\subsection {Clustering Method}
\label{sec:method}

Our clustering method compares the time series by obtaining a dissimilarity matrix based on the scores (obtained as described in Section \ref{sec:PCA}). However, in order to calculate the dissimilarity matrix, we need a method to decide how many principal components to retain and also a suitable measure of distance. Furthermore, once we have obtained the dissimilarity matrix, we will also need to decide which clustering algorithm to use.

  \subsubsection{Distance measures}
  \label{sec:dist measures}

  The success of any clustering algorithm depends on the adopted dissimilarity measure. In this section, we will describe four possible distance measures and discuss their advantages and disadvantages. The proposed distance measures consist of (adaptations and developments of) those adopted in the work reviewed in Section \ref{sec:review}. In our simulation studies in Section \ref{sec:sims}, we will compare the performance of the different clustering algorithms outlined in Section \ref{sec:review}, along with the different possible distance measures.

  The simplest choice for the dissimilarity measure is the squared quadratic distance. This distance measure is adopted by \cite{fryzlewicz2009consistent} (in their paper which classifies nonstationary time series). The advantages of this measure are that it offers good practical performance and is straightforward (and easy) to compute.

  The distance between two time series, $\{X_{t,T}^{(i)}\}_{t=0}^{T-1}$ and $\{X_{t,T}^{(j)}\}_{t=0}^{T-1}$, is the sum of the squared differences between the scores relating to the principal components retained:
  \begin{equation}
   \label{SQD}
   SQD(X_{t,T}^{(i)}, X_{t,T}^{(j)}) = \sum_{k=1}^{p} \Big[\text{Score}_{k}^{(i)}-\text{Score}_{k}^{(j)}\Big]^2,
   \end{equation}
  where $\text{Score}_{k}^{(i)}$ denotes the score (relating to principal component $k$) of time series $X_{t,T}^{(i)}$.
  The value $SQD(i, j)$ is the $(i, j)$th entry of the dissimilarity matrix, $D$.

We propose to develop this simplistic measure by aggregating the scores in the most significant $p$ directions using a \textit{weighted} combination with weights given by the squared singular values. We refer to this measure as the Weighted Squared Quadratic Distance. Therefore, the distance between two time series, $\{X_{t,T}^{(i)}\}_{t=0}^{T-1}$ and $\{X_{t,T}^{(j)}\}_{t=0}^{T-1}$, is the weighted sum of the squared differences between their scores in $p$ directions. Formally:
    \begin{equation}
     \label{WSQD}
     WSQD(X_{t,T}^{(i)}, X_{t,T}^{(j)}) = \frac{\sum_{k=1}^{p} \lambda_k^2 \Big[\text{Score}_{k}^{(i)}-\text{Score}_{k}^{(j)}\Big]^2}{\sum_{k=1}^p \lambda_k^2},
     \end{equation}
    where $\text{Score}_{k}^{(i)}$ is as in equation \eqref{SQD} and $\lambda_k$ denotes the corresponding $k^{th}$ singular value.
    The value $WSQD(i, j)$ is the $(i, j)$th entry of the dissimilarity matrix, $D$.

  Another choice for the dissimilarity measure is outlined in the method by \cite{antoniadis2013clustering}. Their approach applies a singular value decomposition on the covariance matrix between each pair of wavelet transforms to obtain the leading patterns. The distance between two time series is measured by comparing a given number of the leading patterns obtained by the MCA. In particular, they compare the evolutions in time of each pair of leading patterns by measuring how dissimilar their shape is. Therefore, for the $k^{th}$ pair of leading patterns, they take the first derivative of the difference between them. This quantity is bigger (in absolute value) if the two leading patterns evolve very differently through time. Thus, formally, the dissimilarity between two leading patterns, $P_k^{(i)}$ and $P_k^{(j)}$, is measured by:

  \begin{equation}
  d_k(i, j) = |\Delta (P_k^{(i)} - P_k^{(j)})|,
  \end{equation}
  where $\Delta$ represents the first derivative. Finally, \cite{antoniadis2013clustering} aggregate the leading patterns in the most significant $p$ directions using a weighted combination with weights given by the squared singular values:
  \begin{equation}
  D(i, j) = \frac{\sum_{k=1}^p \lambda_k^2 d_k^2(P_k^{(i)}, P_k^{(j)})}{\sum_{k=1}^p \lambda_k^2}.
  \end{equation}

  \cite{rouyer2008analysing} use the following distance (RD) measure adapted from \cite{keogh1998enhanced}:
    \begin{equation}
    \label{RD}
    RD(P_k^{(i)}, P_k^{(j)}) = \sum_{t=1}^{T-1} \tan^{-1}[|(P_k^{(i)}(t)- P_k^{(j)}(t)) - (P_k^{(i)}(t+1) -P_k^{(j)}(t+1))|],
    \end{equation}
    with $T$ being the length of the vectors and $P_k^{(i)}$ and $P_k^{(j)}$ the $k^{th}$ pair of leading patterns for time series $X_{t,T}^{(i)}$ and $X_{t,T}^{(j)}$ respectively. This metric compares two vectors by measuring the angle between each pair of corresponding segments (a segment is defined as a pair of consecutive points of a vector) and is a method for measuring parallelism between curves. The overall distance is then computed as a weighted mean of the distance for each of the $p$ pairs of leading patterns and singular vectors retained (with the weights being equal to the amount of covariance explained by each axis). For the comparison of the time series $X_{t,T}^{(i)}$ and $X_{t,T}^{(j)}$, we compute the distance $DT(i,j)$ according to the following formula:

  \begin{equation}
  DT(i,j) = \frac{\sum_{k=1}^{p}\lambda_k^2(RD(P_k^{(i)}, P_k^{(j)})+RD(\phi_k, \psi_k))}{\sum_{j=1}^{p}\lambda_k^2},
  \end{equation}
 where $\phi_k$ and $\psi_k$  are the $k^{th}$ singular vectors of $X_{t,T}^{(i)}$ and $X_{t,T}^{(j)}$ respectively.

   \subsubsection {Determining the number of principal components to retain}
   \label{sec:numb EOFs}
      In each of the distance metrics we have discussed, we must decide how many axes, $p$, to retain. \cite{antoniadis2013clustering} and \cite{rouyer2008analysing} both decide to use the number of axes that correspond to a fixed percentage of the total covariance (as is common in PCA). However, another method is to select the number of components based on a screeplot. This displays the proportion of variance explained by the (ordered) eigenvalues. $p$ is then selected by looking for an elbow in the screeplot.

      \cite{cho2013modeling} propose selecting this value based on the dimension of the correlation between two curves, $r$. They show retaining $r$ principal components gives a good approximation and also provide a method to estimate the correlation dimension using an information criterion. We do not adopt this method in this work: we obtained similar results with the methods outlined above with less computational burden. However, this may not be the case in other applications of our proposed clustering methodology.

  \subsubsection{Choice of the clustering algorithm}

   Once we have obtained the dissimilarity matrix, we need to decide which clustering algorithm to use. We perform a partitioning around medoids (PAM). This technique admits a general dissimilarity matrix as input and is known to be more robust than other alternatives such as k-means \citep{antoniadis2013clustering}.

  \subsection{Proposed Clustering Method}

   We will now summarise our proposed method of clustering the data.

    \begin{enumerate}
    \item As in Equation \ref{eq:datamat}, obtain an $N \times n$ data matrix, $Q$, (where $n= J \times T$) where the $i^{th}$ row of $Q$ represents the estimated spectra of time series $\{X_{t,T}^{(i)}\}_{t=0}^{T-1}$.
    \item Obtain the scores of each time series as outlined in Section \ref{sec:PCA}.
    \item Retain only $D$ principal components using one of the methods in Section \ref{sec:numb EOFs}.
    \item Form a dissimilarity matrix where the $(i,j)$th entry is the distance (calculated using one of the possible distances defined in Section \ref{sec:dist measures}) between the scores of the $i$-th  and $j$-th time series.
    \item This distance matrix is then used as the input of a clustering algorithm.
    \end{enumerate}

\subsection{Simulation Study}
\label{sec:sims}

In Section \ref{sec:method}, we described our procedure to cluster nonstationary time series. To assess the comparative performance of our proposed procedure with other methods, we conducted a simulation study. We describe the execution and results of this study below.

Firstly, we assumed that each time series was a realisation from one of two groups. Each group has a different EWS, which we used to simulate our data. A data set of $N=100$ (50 simulations from each of the two groups) was generated. For each of the above methods, we obtained a dissimilarity matrix which was the input of a PAM algorithm which clustered the data into two groups. We then compared the clusters with the known group memberships and counted how many time series were correctly clustered (as a percentage). The above procedure was then repeated 100 times and the results for each method were averaged.

\begin{figure}
\centering
\includegraphics[width=0.7\linewidth]{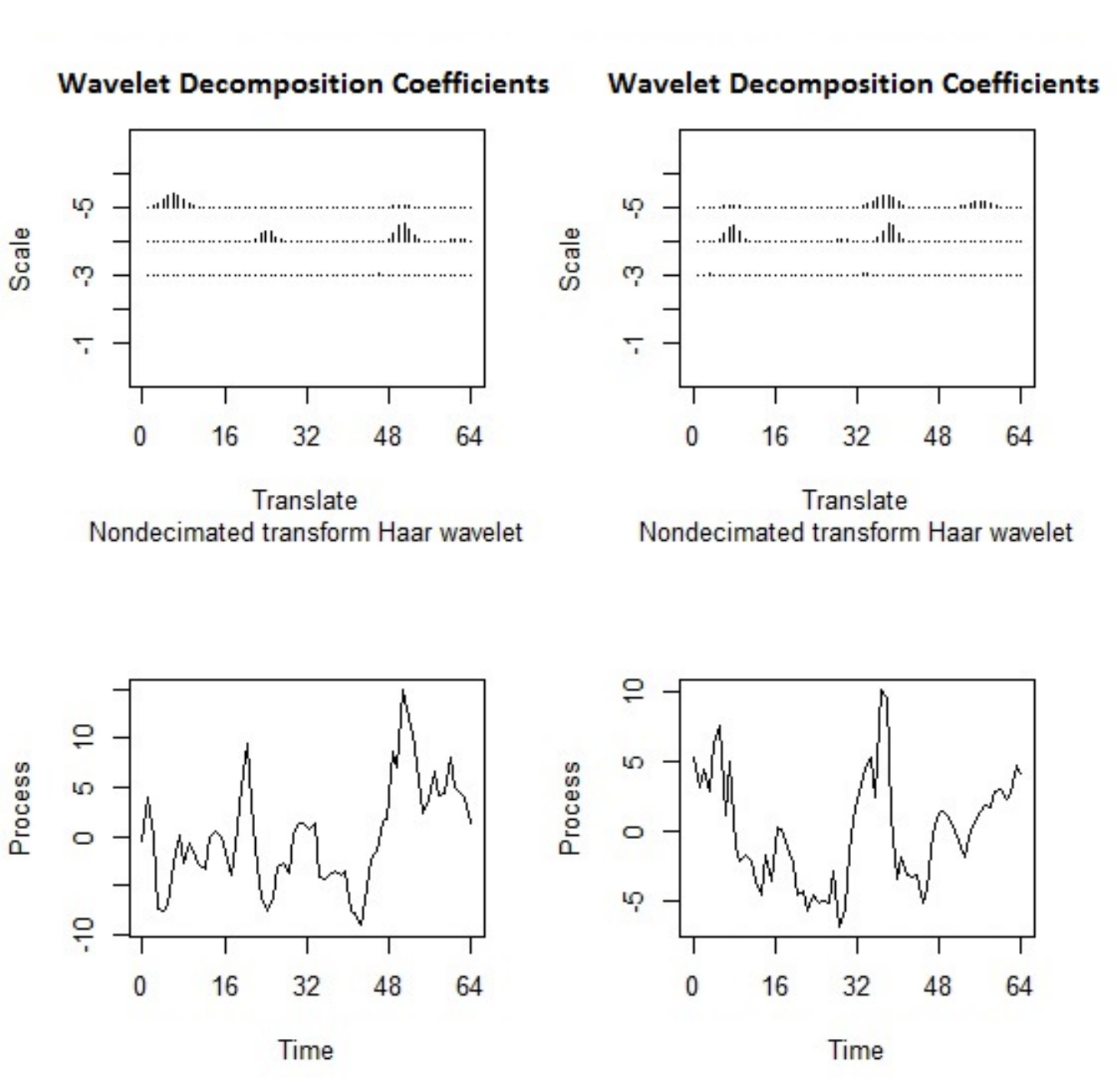}
\caption{Wavelet spectra and realisations from the two groups in the simulation study (based on the circadian dataset).}
\label{fig:specsandts}
\end{figure}

To illustrate the effectiveness of our approach for our application to the circadian dataset, we conducted a simulation study by creating a large sample of synthetic signals from two groups. We simulate the $i^{th}$ time series  of group $g$, from the wavelet spectrum $S_j^{(g)}(z)$, where the spectrum for each of the groups is constructed to display differences which are of interest to the circadian biologist. For example, changes of amplitude and period. Figure \ref{fig:specsandts} shows the wavelet spectra and an example of a realisation from each of the two groups. In particular, these spectra are generated by taking the average spectra of two groups of the observed circadian data set (the 0.5mM and 10mM groups). Therefore, the time series are of length $T = 64$ as this is the length of our truncated circadian dataset.

For this simulation study, we used Daubechies' extremal phase wavelet number 1 both to generate the data and within our clustering methods. The wavelet analysis was performed in the \verb|locits| package in R and the PAM within the \verb|cluster| package. Each periodogram was level smoothed by log transform, followed by translation invariant global universal thresholding and then the inverse transform is applied. For each scale of the wavelet periodogram, only levels 3 and finer are thresholded.

 Firstly, we propose that the similarity of the time series should be based on certain characteristics of the data rather than the raw data itself. Therefore, we propose that clustering time series based on their time-frequency decompositions should give better (and more meaningful) results than using the raw time series. Therefore we begin by reporting the results of clustering based on: the raw data; the wavelet coefficients and the raw wavelet periodogram in Table \ref{tab:raw}. To obtain these results, we clustered the synthetic signals using PAM with the Euclidean distance and the inputs outlined previously. This simulation study shows that clustering based on the raw data and the raw wavelet transform gave poor results (54\% correctly clustered) which supports the assertion that (for this application) clustering based on the second-moment information is preferable. This supports the motivation behind the methods outlined in Section \ref{sec:review}. However, in this paper, we propose  that combining the use of the wavelets with the rigorous stochastic nonstationary time series modelling that is achieved through the assumption of the LSW model, leads to superior methods than those based only on the use of wavelets. Therefore, we also report the results of clustering based on the raw corrected and smoothed wavelet periodogram in Table \ref{tab:raw}. Therefore, this simulation study provides strong evidence that (when clustering based on the second-moment information is preferable and the LSW model is appropriate) methods based on the corrected and smoothed wavelet periodogram (obtained through the assumption of the LSW modelling framework) give significantly better results than methods based on the raw wavelet periodogram. Furthermore, we can also see that using the functional principal components analysis also improves the method from $71\%$ correctly clustered to $75\%$ (see Table \ref{tab:PCs}).

   \begin{table}
   \begin{tabular}{|c|c|c|c|c|}
   \hline Input & Data & Wavelet Transform & Wavelet Periodogram & Corrected Wavelet Periodogram \\
   \hline Correctly Clustered (\%) & 54\% & 54\% & 56\% & 71\% \\
   \hline
   \end{tabular}
   \caption{Simulation study clustering results using PAM with the Euclidean distance and the following inputs: the raw data, wavelet transform, wavelet periodogram and the corrected wavelet periodogram.}
   \label{tab:raw}
   \end{table}

     \begin{table}
     \begin{tabular}{|c|c|c|}
     \hline Method to choose number of EOFs & 90\% of total covariance & Screeplot \\
     \hline Squared Quadratic Distance & 73\% & 75\% \\
     \hline Weighted Squared Quadratic Distance & 77 \% & 78\% \\
     \hline \cite{rouyer2008analysing} Distance & 54\% & 75\% \\
     \hline
     \end{tabular}
     \caption{Comparing methods to select number of principal components for proposed LSW clustering method in the simulation study. Percentages show correct clustering rates.}
     \label{tab:PCs}
     \end{table}

 To examine the effect of the choice of distance measure on our proposed clustering method, we performed the simulation study as above using all four distance measures outlined in Section \ref{sec:dist measures}. The results are summarised in Table \ref{tab:dists}. We can see that, for this simulation study, our method is fairly robust to the choice of distance measure. However, it would seem that the weighted squared quadratic distance gives the best results with $78\%$ of the time series correctly clustered.

  \begin{table}
  \begin{center}
  \begin{tabular}{|c|c|c|c|c|}
  \hline Distance Measure & Squared Quadratic & Weighted Squared Quadratic & \cite{rouyer2008analysing} & \cite{antoniadis2013clustering} \\
  \hline Correctly Clustered & 75\% & 78\% & 75\% & 75\% \\
  \hline
  \end{tabular}
   \end{center}
  \caption{Comparing distance measures for proposed LSW clustering method in the simulation study.}
  \label{tab:dists}
   \end{table}

 We also examined the different methods outlined in Section \ref{sec:numb EOFs} to select the number of principal components to retain for our LSW clustering method. Therefore, we performed our proposed method retaining only two principal components (based on examining the scree plot) and compared this with the situation where we retain the minimal number of components that correspond to $90\%$ of the total covariance. The results are summarised in Table \ref{tab:PCs}. Once again we can see that the LSW clustering method is fairly robust to the way in which we choose the number of principal components to retain. For the squared quadratic and weighted squared quadratic distances there appears to be very little difference between the two approaches. However, the \cite{rouyer2008analysing} distance gives a much lower result when the number of principal components is chosen to explain $90\%$ of the total covariance. On average this method usually uses thirteen principal components whereas the screeplot typically indicates two should be used. Therefore, Table \ref{tab:PCs} suggests that, for this simulation study, the \cite{rouyer2008analysing} distance is much more sensitive to the scores corresponding to relatively smaller proportions of the covariance. Alternatively, \cite{holan2010modeling} argue that there is no reason that the scores associated with principal components that account for more variance should be the important scores in terms of discriminating between the two types of synthetic signal. Therefore, the scores that account for relatively large amounts of the covariance (but are perhaps not chosen by the screeplot approach) could actually not be effective in discriminating between the groups yet are being weighted highly within the distance measure.

 Finally, we compare the LSW method with the methods outlined in Section \ref{sec:review} proposed by \cite{rouyer2008analysing} and by \cite{antoniadis2013clustering}. Both of these benchmark methods do well in practice and represent the state-of-the-art among procedures for clustering nonstationary time series. The results are summarised in Table \ref{tab:methods}. This simulation study provides empirical evidence that our proposed LSW method works very well and outperforms the state-of-the-art among procedures for clustering nonstationary time series. Again we see that (for this particular application) methods based on the second-order information (our LSW method and the \cite{rouyer2008analysing} method) perform better than the method based on the wavelet transform (the \cite{antoniadis2013clustering} method). Moreover, our method, which utilises an unbiased, consistent estimator of the EWS, performs considerably better than the method which uses the raw wavelet periodogram. Finally, \cite{rouyer2008analysing} and \cite{antoniadis2013clustering} use MCA to compare the wavelet representations of two time series at a time and repeat this process for each pair of time series. This simulation study also shows that the method we developed based on \cite{holan2010modeling}, which treats the spectrum as an `image' and performs a PCA on the estimated spectral coefficients of the entire dataset, outperforms the pairwise methods of \cite{rouyer2008analysing} and \cite{antoniadis2013clustering} and is also far less computationally expensive.

  \begin{table}
  \begin{tabular}{|c|c|c|c|}
  \hline Method & \cite{rouyer2008analysing} & \cite{antoniadis2013clustering} & LSW Method \\
  \hline Percentage Correctly Clustered & 65\% & 63\% & 78\% \\
  \hline
  \end{tabular}
  \caption{Comparing the proposed LSW clustering method with the methods outlined in \cite{rouyer2008analysing} and \cite{antoniadis2013clustering} in the simulation study.}
  \label{tab:methods}
  \end{table}

\section{Circadian Data Results}\label{sec:app}
In this section, we apply the clustering method developed in Section \ref{sec:method} to the circadian data that motivated this work.

\subsection{Pre-processing the Data}
As stated in Section \ref{sec:model}, the time series should be a zero mean process. Therefore, we estimate the mean \textit{of each series} and subtract this from each series. Figure \ref{fig:Circgroups} shows each realisation from each group (in grey) along with the group average (in red) for our truncated zero-mean dataset.

\subsection{Preliminary Analysis}
\label{sec:prelim}

For each plant we calculated the corrected wavelet periodogram estimate of the EWS. The wavelet analysis is implemented in the \verb|locits| package in R (available from the CRAN package repository). For this analysis we used Daubechies' extremal phase wavelet number 1. Each periodogram was level smoothed by log transform, followed by translation invariant global universal thresholding and then the inverse transform is applied. For each scale of the wavelet periodogram, only levels 3 and finer are thresholded.

\subsection{Clustering Results}

\begin{figure}
\centering
\includegraphics[width=0.7\linewidth]{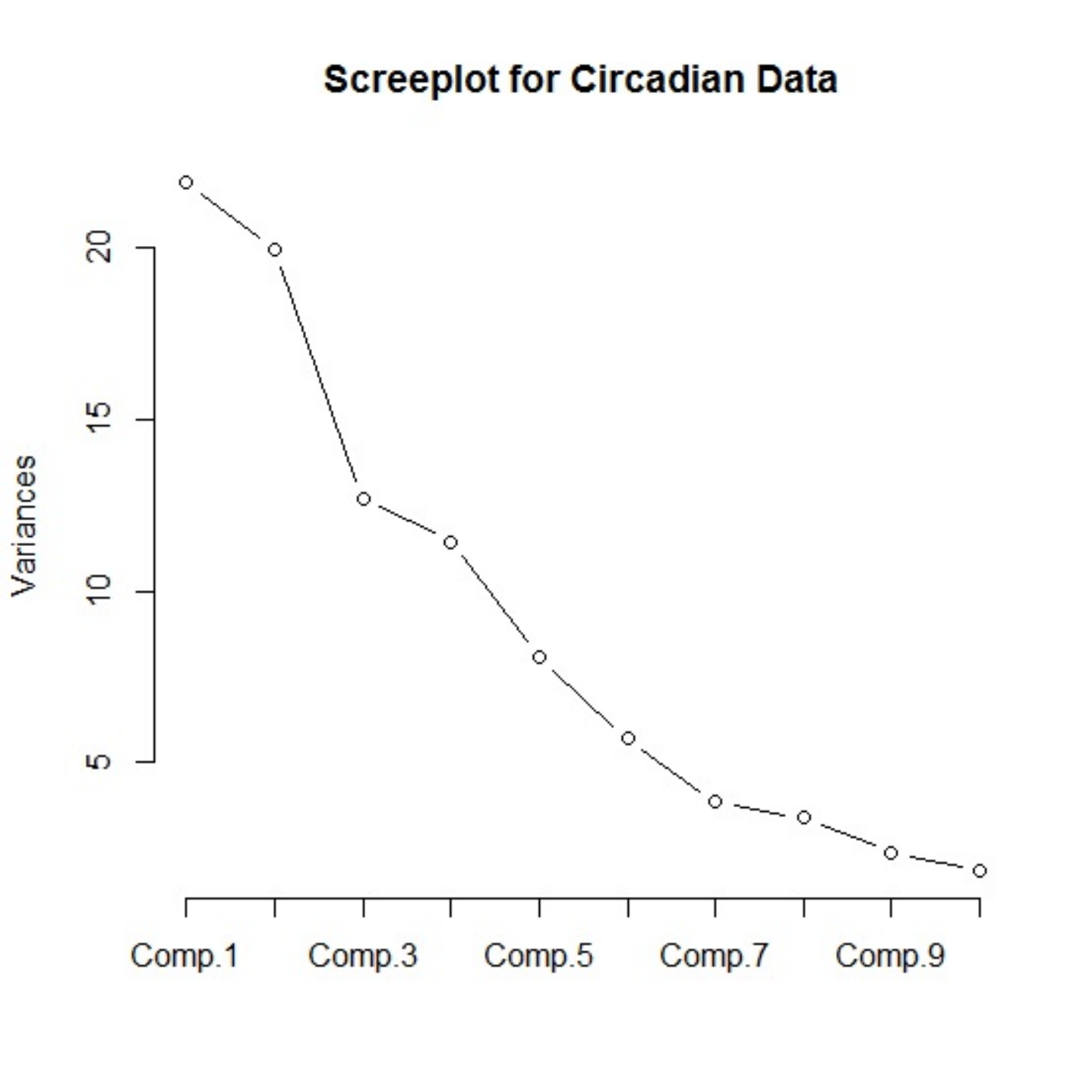}
\caption{The screeplot for the circadian data.}
\label{fig:Circscree}
\end{figure}

\begin{figure}
\centering
\includegraphics[width=0.7\linewidth]{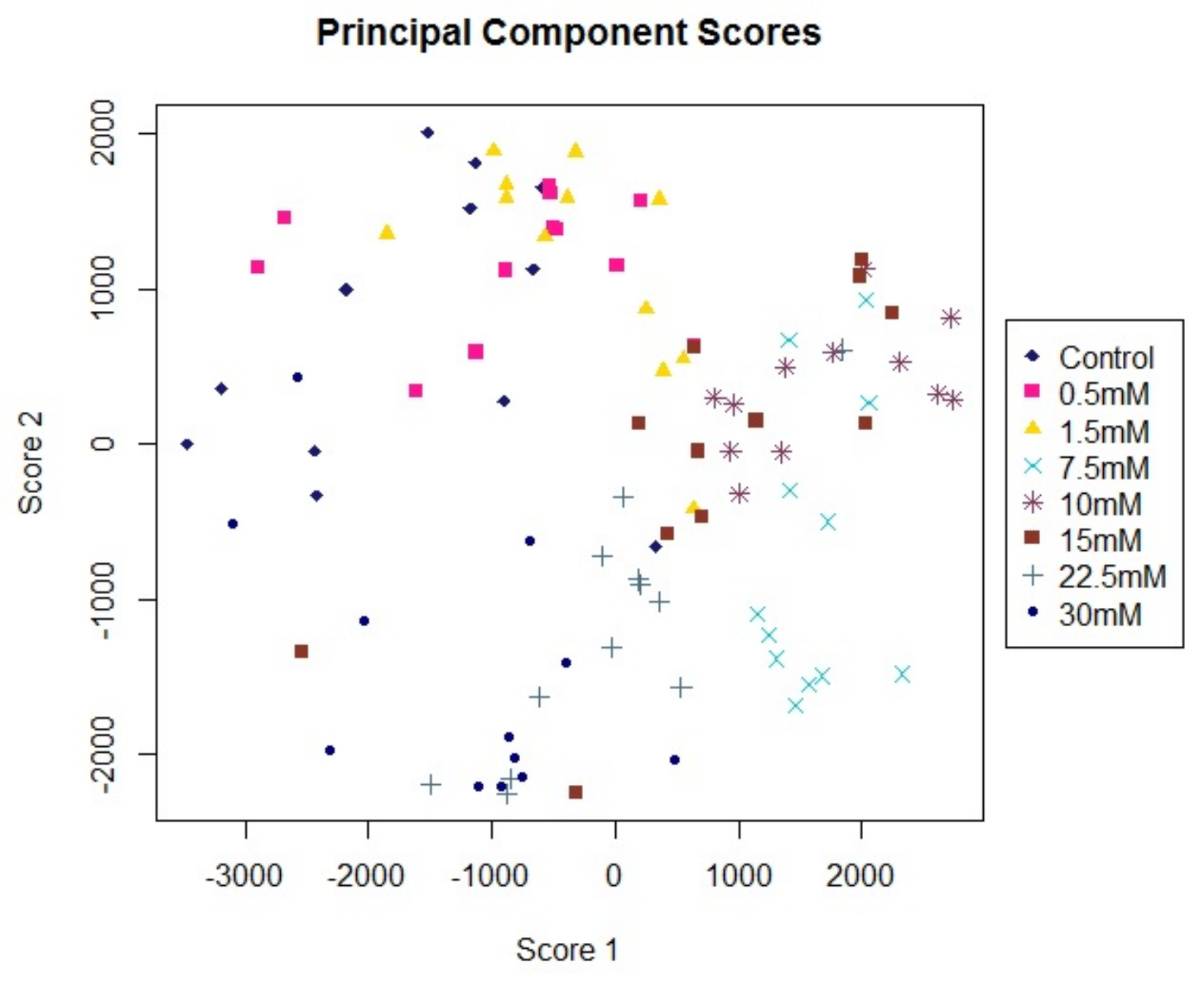}
\caption{The circadian data projected onto the first two principal components obtained from the LSW clustering method.}
\label{fig:Circscoresall}
\end{figure}

To cluster the data, we decided to retain two principal components based on examining the screeplot in Figure \ref{fig:Circscree}. We can also plot the data in relation to the scores of the first two principal components. In Figure \ref{fig:Circscoresall} we have plotted the data projected onto the first two principal components by group. On examining Figure \ref{fig:Circscoresall} we can see that the lower concentrations occupy the top-left area of the plot.

\begin{figure}
\centering
\includegraphics[width=0.7\linewidth]{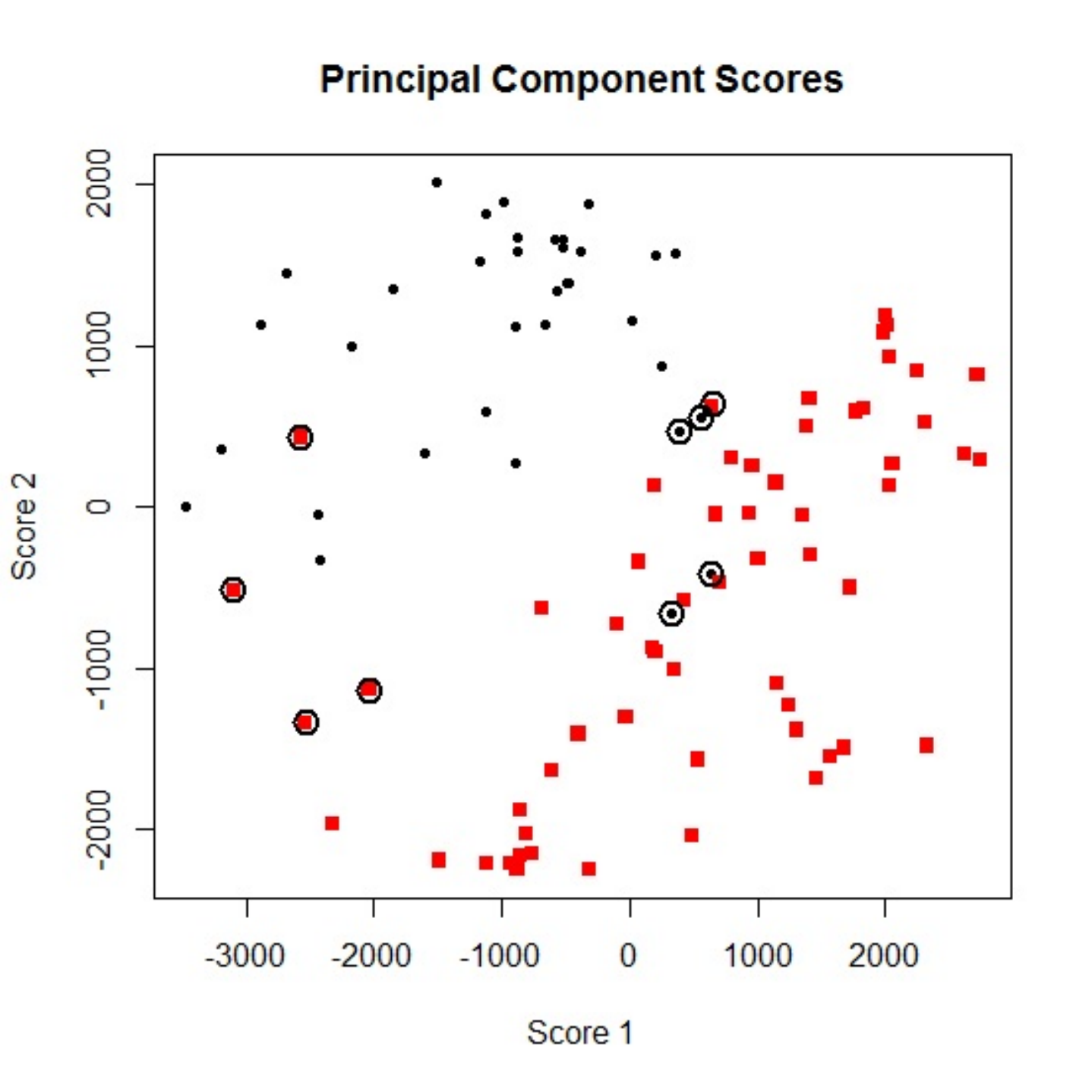}
\caption{The circadian data projected onto the first two principal components obtained from the LSW clustering method. The control and concentrations 0.5mM and 1.5mM are represented by black circles, the higher concentrations by red squares. The circled points highlight plants which were misclustered.}
\label{fig:Circscores}
\end{figure}

% wrong group 1: 6 13 26 34 36 ( 5)

% wrong group 2: 62  85 86  90 (4)

% 2 PCs

\begin{figure}
\centering
\includegraphics[width=0.7\linewidth]{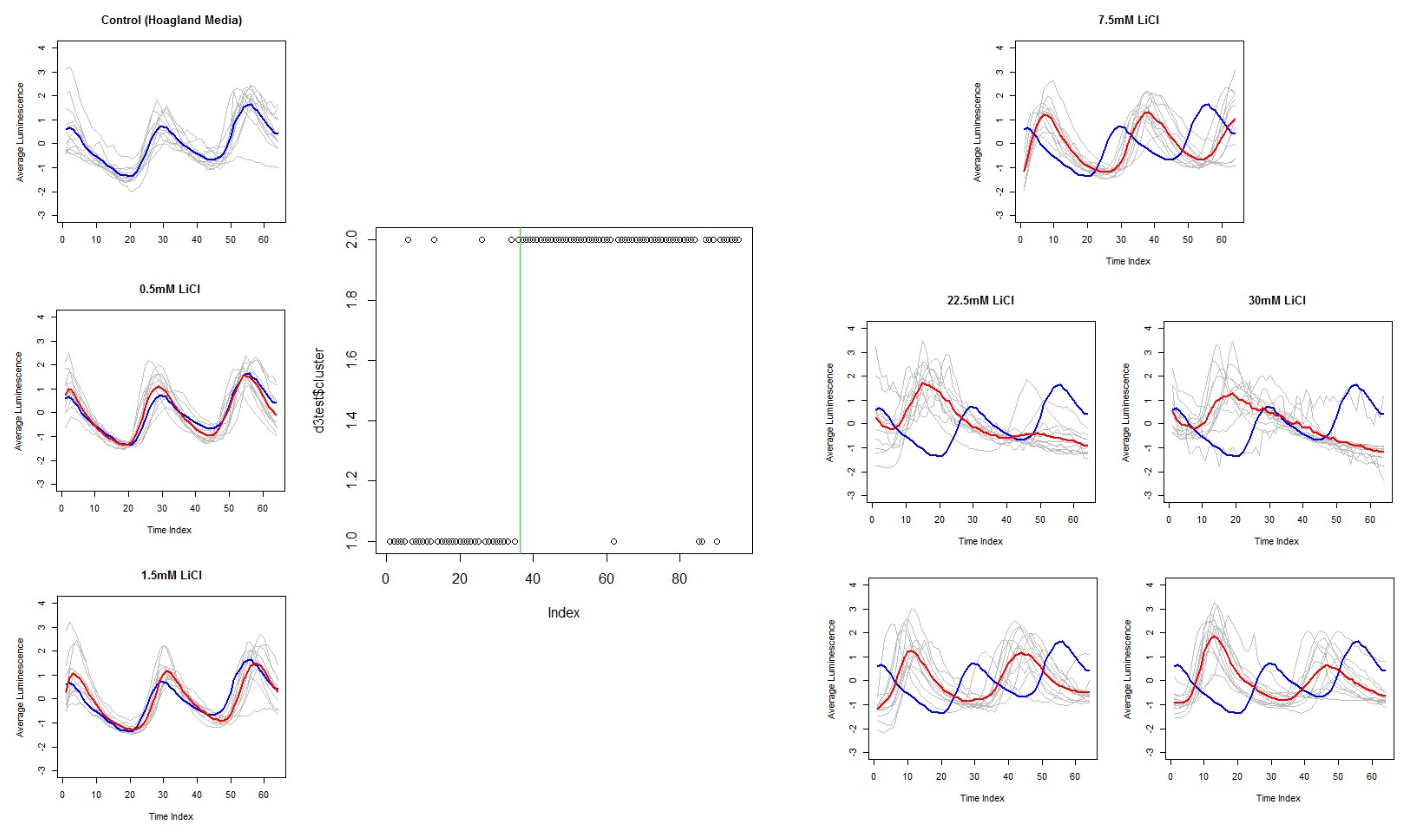}
\caption{The results of clustering the circadian dataset into two groups using the LSW method. The central plot shows the cluster labels for each time series (labelled 1-96). Thus, those with index 1-36 belong to the first 3 lower concentration groups with successive blocks of 12 time series belonging to rising concentrations. For reference, the time series and averages for each of the groups are shown beside the results.}
\label{fig:Circclust2}
\end{figure}

We then obtained a dissimilarity matrix by computing the weighted squared quadratic distance between the first two scores of each time series. We used the "cluster" R package to perform a PAM. The results of clustering the data into two groups using our methods are shown in Figure \ref{fig:Circclust2}. We chose to cluster into two groups since one biological application of this method could be to ascertain at which level of lithium we start to see an effect. Furthermore, we could characterise the effect this exposure is having with the results of this analysis.

\subsection{Discussion of Findings}

On examining Figure \ref{fig:Circclust2}, we can see that this method has effectively sorted the data into two groups:
\begin{enumerate}
\item the control and concentrations 0.5mM and 1.5mM and
\item the higher concentrations.
\end{enumerate}

 This is to be expected as Figure \ref{fig:avplot}shows that these groups display a similar average. Furthermore, we also note that the increase in concentration from 1.5mM to 7.5mM is relatively large. Therefore, our method differentiates between the lower and higher concentrations of lithium. This would suggest that there exists a "threshold" of the amount of lithium the plant can tolerate. Furthermore, this study implies that this threshold is 7.5mM. However, as stated above, there is a relatively large jump between 7.5mM and the 1.5mM. Therefore, this study would imply that more research should be done with finer gradients between concentrations in the range 1.5mM to 7.5mM to find a more precise limit.

\begin{figure}
\centering
\includegraphics[width=0.7\linewidth]{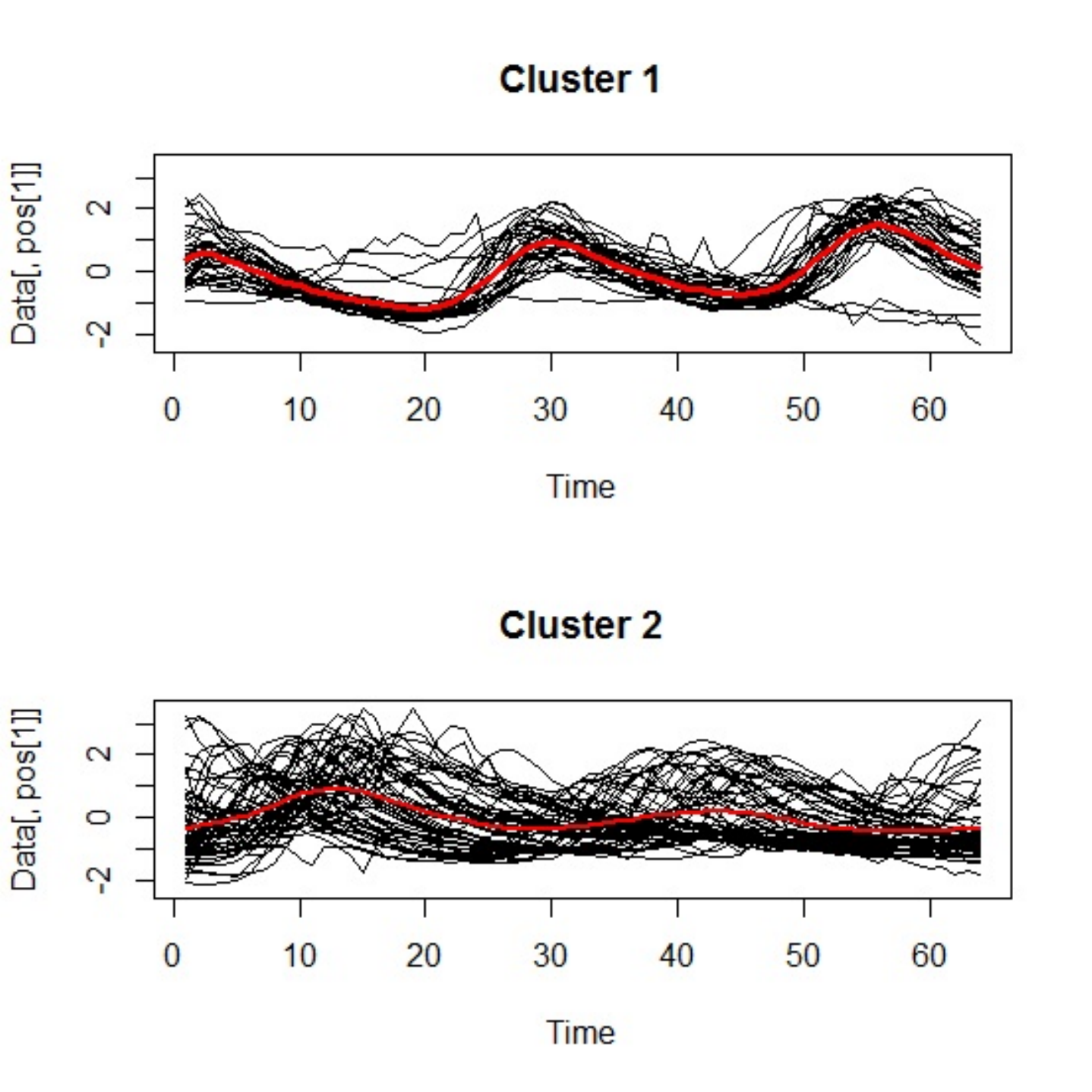}
\caption{The results of clustering the circadian dataset into two groups using the LSW method. The time series of the clustered groups are shown in black along with the cluster average in red.}
\label{fig:Circclustplots}
\end{figure}

\begin{figure}
\centering
\includegraphics[width=0.7\linewidth]{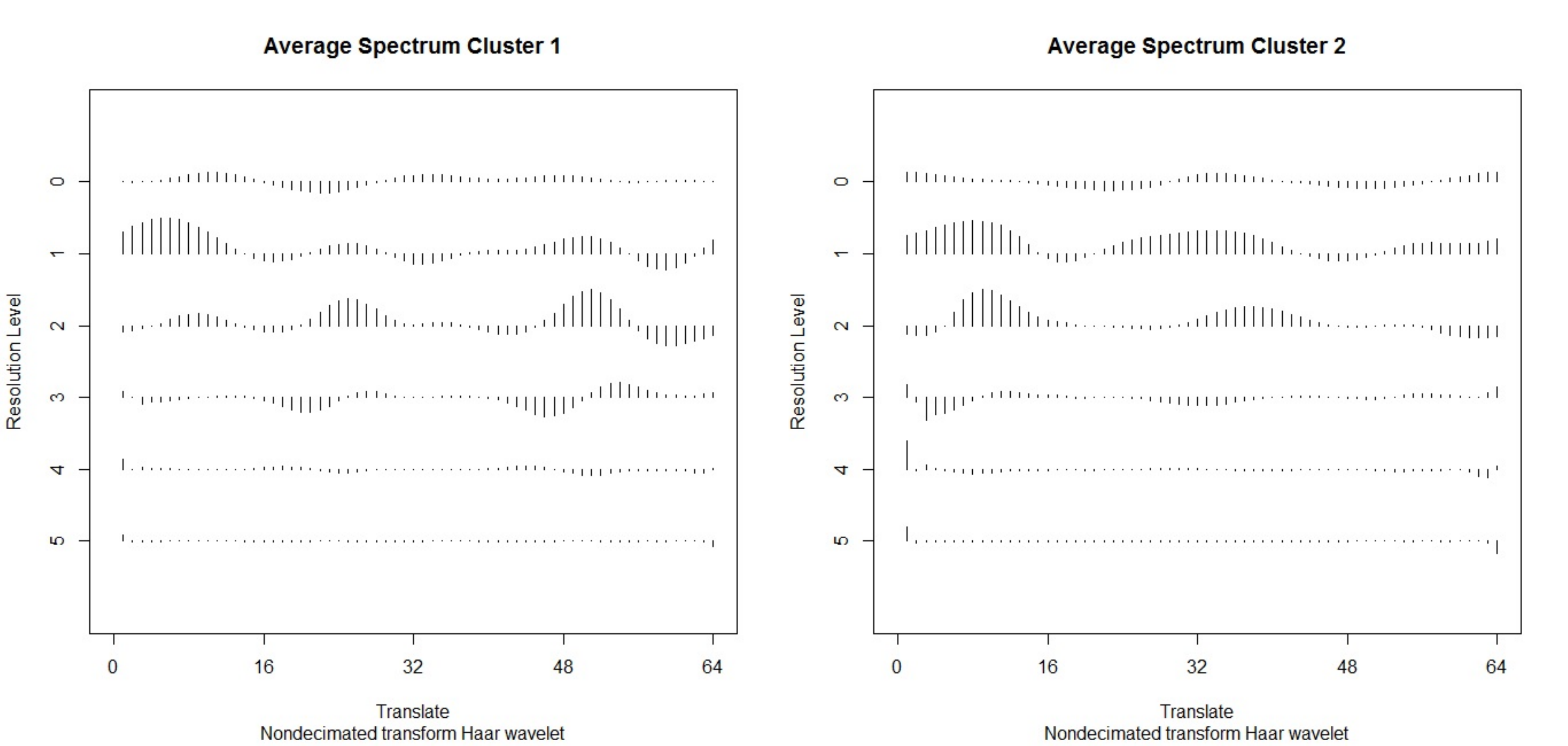}
\caption{The average spectrum for each cluster resulting from clustering the circadian dataset into two groups using the LSW method. Cluster 1 corresponds to the lower concentration of lithium and cluster 2 the higher concentration.}
\label{fig:avcirc}
\end{figure}

The proposed method also allows to characterise these groups. The time series of the clustered groups (in black) along with the cluster average (in red) are shown in Figure \ref{fig:Circclustplots}. We also plot the average spectrum for each cluster in Figure \ref{fig:avcirc}. These figures suggest that the period of all the plants changes (from 24 hours) after exposure to constant light. In particular, the average spectrum of cluster 1 has a peak in resolution level 1 beginning at 0 hours and then two peaks in resolution level 2 at around 20 hours and after 48 hours. This movement through the scales to the finer resolutions as time progresses implies that the period changes with time. This is more evidence to suggest that a single period estimate, the standard practice in the circadian community, does not effectively characterise the data. Furthermore, the large coefficients in resolution level 1 throughout the experiment in the average spectrum of cluster 2 also imply that this group has a longer period than cluster 1. Finally, we propose that exposure to higher concentrations of lithium reduces the amplitude of the signals as time progresses. This is apparent in the average spectrum of cluster 2 (in Figure \ref{fig:avcirc}) as the magnitude of the spectral coefficients decreases as time progresses.

\begin{figure}
\centering
\includegraphics[width=0.7\linewidth]{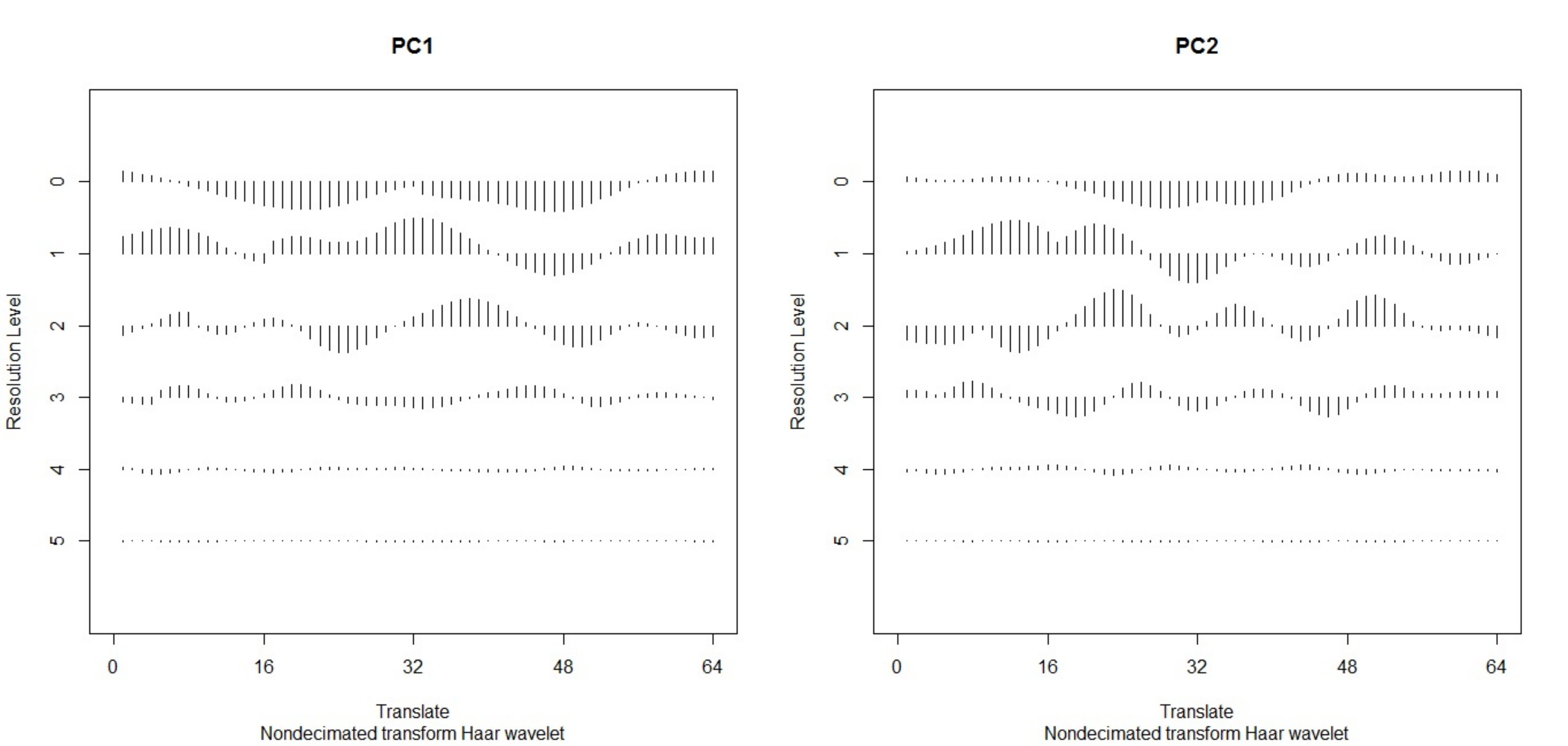}
\caption{Principal components 1 and 2 obtained by clustering the circadian dataset into two groups using the LSW method.}
\label{fig:PCsall2}
\end{figure}

Let us now inspect the first two principal components in Figure \ref{fig:PCsall2}. The first principal component identifies the differences in resolution level 2 of the evolutionary wavelet spectra of the two groups. The peak just after 32 hours corresponds to a peak in the spectrum of cluster 2 and the troughs correspond with the peaks in the spectrum of cluster 1. We can also see a similar relationship in resolution level 1 with peaks in the principal component corresponding to peaks in the spectrum of cluster 2 and troughs corresponding with the peaks in the spectrum of cluster 1. This gives a negative score for cluster 1 which we can see in Figure \ref{fig:Circscores}. Therefore, the first principal component could represent the longer period associated with exposure to lithium.

\begin{figure}
 \centering
 \includegraphics[width=0.7\linewidth]{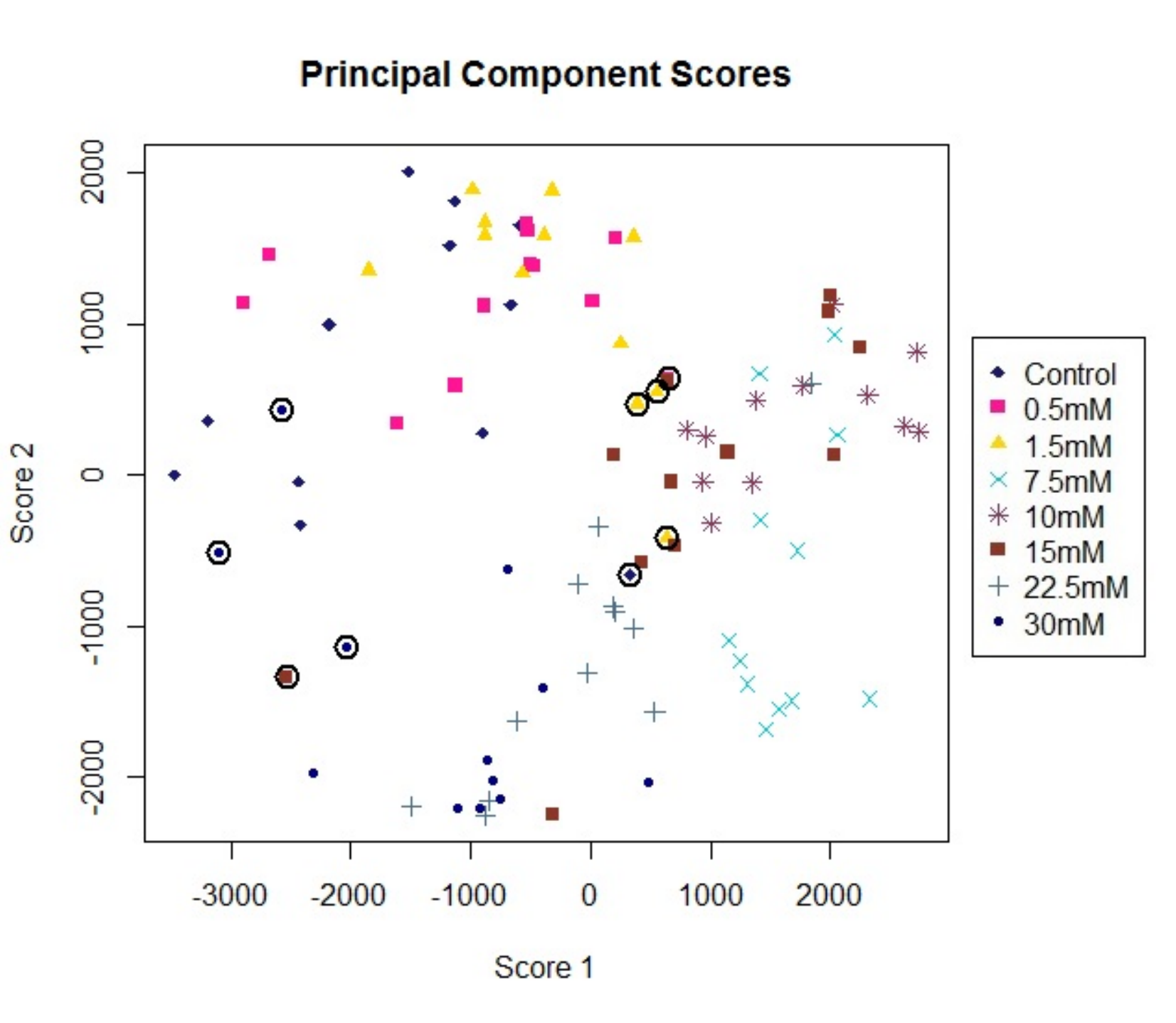}
 \caption{The data projected onto the first two principal components obtained by clustering the circadian dataset into two groups using the LSW method. The circled points represent plants which were miss-clustered (one from the control group; one from 0.5mM,three from 1.5mM; one from 15mM; three from 30mM).}
 \label{fig:Circscoresallwrong}
 \end{figure}

\begin{figure}
  \centering
  \includegraphics[width=0.7\linewidth]{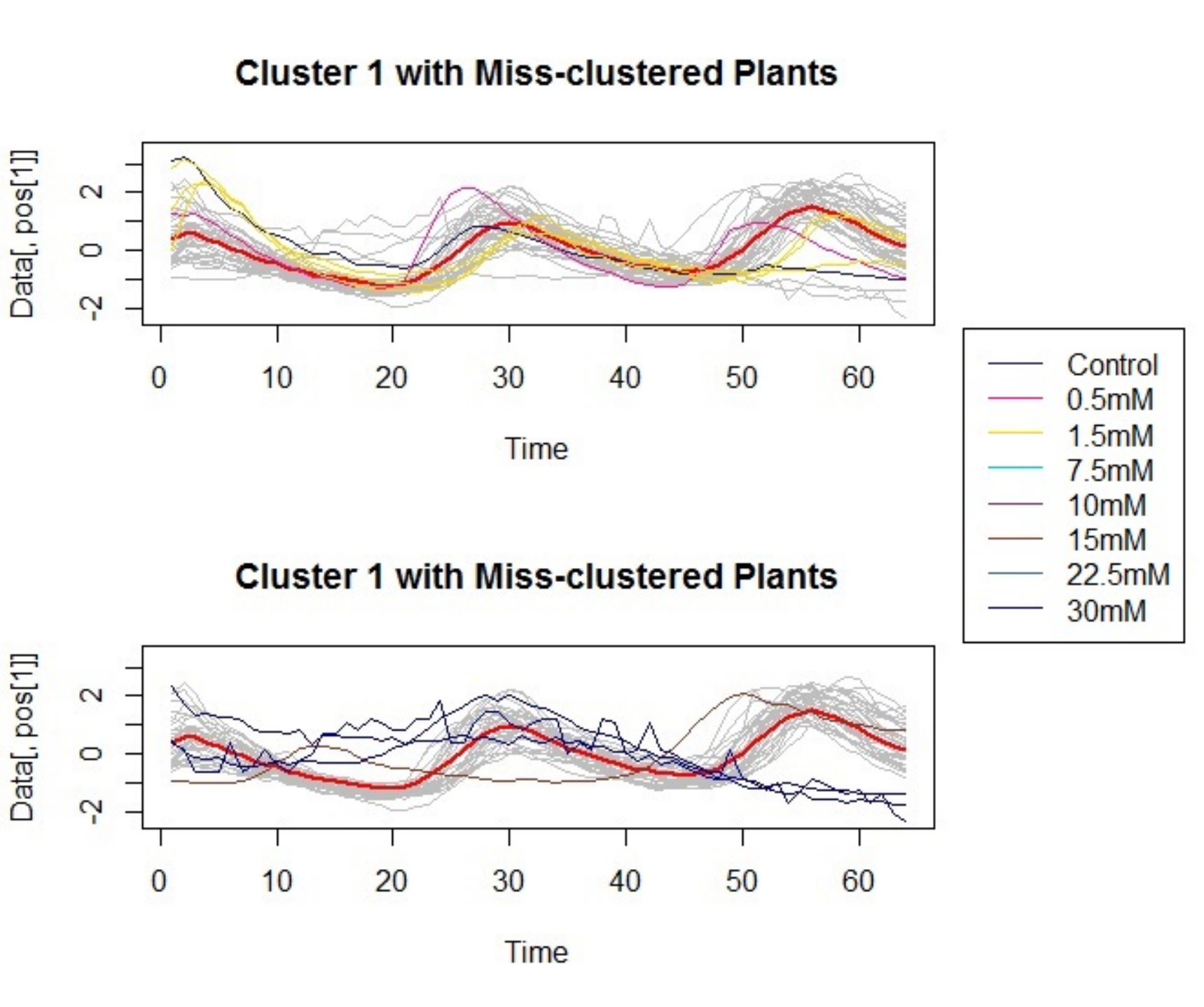}
  \caption{The time series of cluster 1 (in grey) along with the cluster average (in red) obtained by clustering the circadian dataset into two groups using the LSW method. Top panel: the time series from lower concentrations that were clustered into group 2 (one from the control group; one from 0.5mM,three from 1.5mM). Bottom panel: the time series from higher concentrations that were clustered into group 1 (one from 15mM; three from 30mM). In both figures, the colours show the group concentration.}
  \label{fig:Circclustwrong}
  \end{figure}

 On interrogating Figure \ref{fig:Circclust2}, it seems that 5 of the "lower concentration" time series and 4 of the "higher concentration" time series were "incorrectly" clustered. Therefore, we could say that this method has correctly clustered approximately 89\% of the circadian data. The incorrectly clustered points are circled in Figure \ref{fig:Circscoresallwrong}. Furthermore, Figure \ref{fig:Circclustwrong} shows the observed time series of the miss-clustered points in relation to the observations from cluster 1. From the top panel of Figure \ref{fig:Circclustwrong} we can see that the lower concentration time series that were assigned to cluster 2 seem to exhibit the dampening effect and period-lengthening which characterises the exposure to higher concentrations. This could mean that the effects of adding lithium can be seen in plants which are not exposed to higher concentrations of lithium. The lower panel of Figure \ref{fig:Circclustwrong} displays the plants exposed to higher concentrations of lithium which were assigned to cluster 1. These time series do not seem to dampen in the same way as the other higher concentration time series and share a peak at around 30 hours which suggests that this property characterises the control group and lower concentrations. Furthermore, this indicates that some plants may be more resilient to the exposure of lithium. Therefore, we conclude that though there are certain features which characterise exposure to lithium, not all plants react in the same way.

In conclusion, our analysis demonstrated that a threshold exists with regards to the amount of lithium a plant can tolerate before its clock is distinctly affected. Furthermore, we found that this boundary is between 1.5mM and 7.5mM of LiCl. Finally, we were also able to characterise these effects using our proposed analysis methods. Therefore, adding concentrations of lithium above 1.5mM has two main effects (i) period lengthening  and (ii) amplitude dampening.

In the circadian community, such data is currently analysed by simply estimating a (constant) period for each time series (by means of Fourier analysis). Our proposed multiscale analysis however, has clearly demonstrated the unsuitability of this approach. This is due to the underlying nonstationary character of the data coupled with the lithium concentration effects. Our time-scale method provides further insight into this area of study as data is characterised using a time-varying period as opposed to one period estimate.

\section{Conclusions and Further Work}\label{sec:concs}

In this paper we developed a new procedure for clustering circadian plant rhythms by modelling them as nonstationary wavelet processes and exploiting their local time-scale spectral properties. Our method combines the advantage of a wavelet analysis with the benefits of rigorous stochastic nonstationary time series modelling. When compared to competitor (non-model based) methods, we found that the locally stationary wavelet model brought clear gains both for simulated and real data.

The proposed model-based clusterings can be used to produce visualisations helpful in answering questions such as `what other concentrations of lithium produce similar effects in plants?' and `what characterises the different types of reactions present in this dataset?' The answers to these questions have important implications for understanding the mechanism of the plant's circadian clock and also environmental implications associated with soil pollution. We also showed that our method has desirable properties such as low sensitivity to the choice of distance measure and number of principal components to retain. We believe these results show the method's suitability in organising and understanding multiple nonstationary time series such as the gene expression levels in our circadian dataset.

At this point we should note that method is not restricted to the dataset analysed in this paper, but can be applied to other circadian datasets. For example, we could extend this experiment to include the results of exposure to other elements and answer the question `which other elements in the periodic table, and at which concentrations, produce similar kinds of reactions in plants?' We can also extend the dataset to include plants with \emph{deficiencies} of an element. This would also enable deeper understanding of the circadian clock mechanism. It is also possible to observe the expression of other genes in these experiments. Therefore, our methods could also be used to cluster nonstationary time series in order to identify genes potentially involved in, and regulated by, the circadian clock. Alternatively, it is possible to simultaneously measure the expression levels of multiple genes. Therefore, an avenue for further research would be to extend our methods to cluster \emph{multivariate} time series.

In Section \ref{sec:data}, we discussed the propensity of the recording equipment to break down resulting in gaps in the data. Another area of future work is to adapt current methods under the presence of missingness, or `gappy' data, often arising in experimental data. This estimate could then be used as a classification signature or within our clustering procedure.

The wavelet system gives a representation for nonstationary time series under which we estimate the wavelet spectrum and cluster the data based on these coefficients. We have found in simulations that our method is fairly robust to the choice of wavelet. However, it may be that certain wavelets are better suited to modelling and discriminating between certain datasets. Furthermore, it may be that different wavelets identify different features of the data. Thus an interesting area of further work would be to derive a procedure for determining significant data features and discriminating between groups.

\bibliography{Cluster_draft.bbl}

\end{document}